# Boron nitride nanotube precursor formation during high-temperature synthesis: kinetic and thermodynamic modelling

**Yuri Barsukov[1], Omesh Dwivedi[2], Igor Kaganovich[2], Sierra Jubin[2], Alexander Khrabry[2], and Stephane Ethier[2]**

[1] Plasma Physics Department, Peter the Great Saint-Petersburg Polytechnic University, Saint-Petersburg, 195251 Russia
[2] Princeton Plasma Physics Laboratory, Princeton University, Princeton, NJ, 08543 USA

E-mail: barsukov.yuri@gmail.com

## Abstract

We performed integrated modelling of the chemical pathways of formation for boron nitride nanotube (BNNT) precursors during high-temperature synthesis in a $B/N_2$ mixture. Integrated modelling includes quantum chemistry, quantum-classical molecular dynamics, thermodynamic modelling, and kinetic approaches. We demonstrate that BN compounds are formed via the interaction of molecular nitrogen with small boron clusters, rather than through interactions with less reactive liquid boron. (This process can also be described as $N_2$ molecule fixation.) Liquid boron evaporates to produce these boron clusters ($B_m$ with $m \leq 5$), which are subsequently converted into $B_mN_n$ chains. The production of such chains is crucial to the growth of BNNTs because these chains form the building blocks of bigger and longer BN chains and rings, which are in turn the building blocks of fullborenes and BNNTs. Additionally, kinetic modelling revealed that $B_4N_4$ and $B_5N_4$ species in particular play a major role in the $N_2$ molecule fixation process. The formation of these species via reactions with $B_4$ and $B_5$ clusters is not adequately described under the assumption of thermodynamic equilibrium, as is demonstrated in our kinetic modelling. Thus, the accumulation of both $B_4N_4$ and $B_5N_4$ depends on the background gas pressure and the gas cooling rate. Long BN chains and rings, which are precursors of the fullborene and BNNT growth, form via self-assembly of components $B_4N_4$ and $B_5N_4$. Our modelling results – particularly the increased densities of $B_4N_4$ and $B_5N_4$ species at higher gas pressures – explain the experimentally observed effect of gas pressure on the yield of high-quality BNNTs. The catalytic role of hydrogen was also studied; it is shown that HBNH molecules can be the main precursor of BNNT synthesis in the presence of hydrogen.

## 1. Introduction

### 1.1 Literature review

Boron nitride nanotubes (BNNTs) have a structure similar to carbon nanotubes, but with different properties [1] due to the polarity of B-N bonds. In the past, high-quality BNNTs have been synthesized at high temperatures using arc discharges [2,3], laser ablation [4–6], and inductively-coupled plasma (ICP) torches [7–9]; at relatively low temperatures, BNNT synthesis has been performed via ball-milling [10] and chemical vapor deposition [11–13]. High-temperature synthesis by arc discharges or ICP torches involves the injection of $N_2$ gas and B powder into the plasma zone first [7], where the gas temperature is high enough (> 8000K [8]) to completely vaporize the boron powder while the molecular nitrogen stays intact [14]. The gas then flows into the cooler afterglow zone, where synthesis of BNNTs occurs. In this zone, the gas temperature drops precipitously at a rate of about ~$10^5$ K/s [8]. It was experimentally observed that the growth of BNNTs begins at ~3000K [9]; therefore, to accurately predict how BNNTs form in these high temperature synthesis scenarios, we have made an effort to predict the gas composition as the mixture cools down from 8000K to 2000K.

There are many aspects of BNNT growth that are not well understood [15]. Three growth mechanisms have previously been discussed in the literature: root growth [4,8], open-end growth [16], and growth by self-assembly [17].

In the root growth model, BNNTs grow on the surface of boron droplets, and growth occurs at the point of attachment to the droplet (the root). Boron droplets are formed at temperatures (~4100K at 1 atm. [18]) significantly higher than the temperature of BNNT growth, so droplet formation occurs earlier in the gas cooling process. According to the root-growth mechanism, atomic nitrogen, molecular nitrogen, or B-N gas species are dissolved in the droplets, and these molecules feed the growth of BNNTs [9,17,19].

A related mechanism of BNNT formation is open-end growth, in which B-N compounds react with an unattached end of a BNNT. The other end of the nanotube may be attached to a boron droplet, though not necessarily [16].

The third mechanism, growth by self-assembly, was discussed in Ref. [17]. Based on molecular dynamic (MD) simulations, Han and Krstic suggested that BNNT formation can occur without boron droplets via the self-assembly of small species with B-N bonds, such as BN dimers. However, the production of these species with B-N bonds was not investigated. Here, we have focused on precursor formation for this self-assembly mechanism of BNNT growth.

In our early research [14] it was shown that small B-N species such as BN, $B_2N$, and $N_2B$ are generated in boron and nitrogen gas mixtures. However, the thermodynamic equilibrium densities of all small B-N compounds considered in that study were shown to be very low compared to the densities of molecular $N_2$ and liquid boron at temperatures suitable for BNNT growth (~2600K). Considering our earlier research, which demonstrated that $N_2$ molecules do not directly react with liquid boron droplets [19], *it remained unclear how $N_2$ molecules and liquid boron are converted into compounds that contain B-N bonds. The main goal of this study is to understand this crucial step of nitrogen fixation in the BNNT synthesis process.*

## 1.2 Our contribution

In our previous thermodynamic approach [14] we studied the B-N mixture composition at equilibrium and we only considered small species that contain B-N bonds (BN, $B_2N$, $N_2B$). In this study, we significantly expanded the number of B-N species under consideration, adding $B_mN_n$ chains and $B_nN_n$ fullborenes (Section 2.2). We showed that at temperatures near 2600K or slightly higher, particular $B_mN_n$ chain species have the highest densities in the reaction mixture, while at T<2600K fullborenes become the major component. The densities of small B-N gas species, such as BN, $N_2B$ and $B_2N$, have a maximum value at temperatures near the boron condensation point (4100K at 1 atm) and drop quickly as the gas cools. This contradicts previous assumptions that these species act as direct precursors for BNNT growth [14,17,19]. *In summary, the densities of the small B-N species are very low near the BNNT growth temperature (~2600K), but the densities of $B_mN_n$ chains at this temperature are high. Therefore, we conclude that the latter are precursors in the formation of BNNTs.*

In addition, we show that these $B_mN_n$ chains are synthesized via reactions between molecular nitrogen and small boron clusters ($B_m$ with $m \leq 5$) at temperatures ranging from the temperature of BNNT growth (~2600K) to the boron condensation point (4100K at 1 atm.). This process is the mechanism for $N_2$ fixation during high temperature BNNT synthesis. Liquid boron droplets act as a source of $B_m$ clusters in the gas phase, but do not themselves fix nitrogen molecules. Small boron clusters in gas phase should have a higher reactivity than liquid boron, and smaller clusters should have a higher reactivity than larger clusters [20]. (For example, the reactivity of $B_7^+$ clusters with $N_2O$ is much smaller compared to that of $B_6^+$ clusters.) This is due to the fact that larger boron clusters are more coordinatively saturated (have fewer dangling bonds). Further research is needed to systematically study the transformation of $B_mN_n$ chains into BNNTs, and to investigate additional details regarding the role of boron droplets in this transformation. (See the Discussion section.)

We also report that the effectiveness of the fixation of $N_2$ depends on the process conditions. *Namely, it is shown that boron consumption and $B_mN_n$ chain generation can be enhanced by increasing the background gas pressure, by slowing the gas cooling rate, and by decreasing the initial boron fraction in the reaction mixture.*

## 1.3 Organization of the paper

The organization of the paper is as follows: details of the computational methods used and the results of modeling are discussed in Section 2. This includes an investigation into the binding energy and thermodynamic stability of $B_mN_n$ clusters by calculation of their Gibbs free energies in Section 2.1, global modelling of the equilibrium mixture composition in Section 2.2, quantum chemistry studies of the reaction pathways for $N_2$ fixation in Section 2.3, global kinetic modelling of the $N_2$ fixation process with time-dependent reaction rates in Section 2.4, and finally quantum-classical molecular dynamics simulations of small $B_mN_n$ chains aggregating into larger $B_mN_n$ chains in Section 2.5. Discussion of the results and conclusions are addressed in Sections 3 and 4, respectively.

## 2 Computational results

Here we describe the multiple approaches used to model BNNT precursor formation. First, we determined the composition of the gas mixture as a function of temperature for pressures of 1 atm and higher under the assumption of thermodynamic equilibrium. This method is referred to throughout the paper as the ‘thermodynamic approach.’ Next, chemical kinetic calculations were carried out to check whether or not the assumption of thermodynamic equilibrium is valid during typical gas cooling conditions for BNNT synthesis.

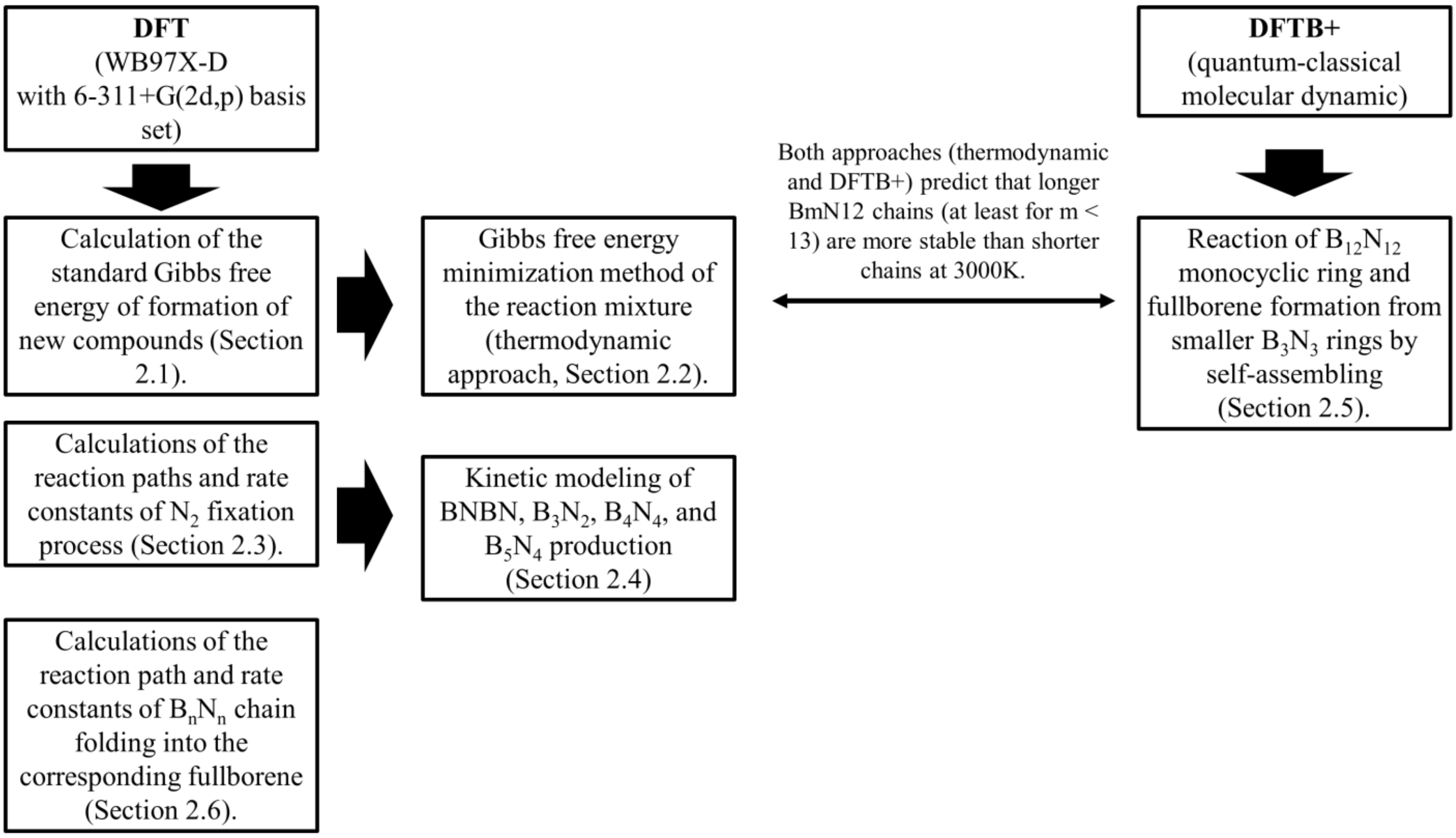

**Figure 1.** Schematic diagram describing computational techniques which were used in our research.

Thermodynamic and kinetic modelling requires the calculation of chemical potentials (the standard Gibbs free energy $\Delta_f G^0$) and rate constants for each relevant reaction. These were obtained through DFT (density functional theory) quantum chemistry (QC) calculations using the GAMESS software package [21]. We used an unrestricted WB97X-D [22,23] hybrid functional and the 6-311+G(2dp) basis set. For large $B_nN_n$ clusters, where n > 13, the Gaussian-16 software package [24] and 6-31+G(d) basis set were used.

Reaction paths, some of which are described in Section 2.3, were investigated using DFT methods, and in some cases MCSCF (multi-configuration self-consistent field) methods [25] were also used to verify the DFT approach. The MCSCF method includes single- and double-excitations (MCSCF-SD) with 30776 configuration state functions in the active space. The geometries of all considered structures were first optimized using DFT and then, if applicable, further optimized using MCSCF-SD, at which point single point energy calculations were carried out using MCSCF with single-, double- and triple-excitations (MCSCF-SDT) without further optimization. All transition states (TSs) were identified by the presence of one imaginary frequency in the Hessian matrix and by an IRC method (intrinsic reaction coordinate).

Reaction paths for molecules larger than BNBN were optimized using the DFT method only, as MCSCF calculations are more computationally intensive. Further details on the applicability and accuracy of DFT approaches are discussed in the supplementary material, on an example of $BN_2$ molecule formation [26].

The calculated rate constants $k_i$ of the gas phase reactions, which are used in the kinetic model (Section 2.4), can be found in the supplementary materials. All rate constants are presented in the Arrhenius form:

$$k_i = A_i * (\mathrm{T}/298.15)^n * \exp\left(-\frac{E_a(i)}{k_B T}\right). \qquad (1)$$

The reaction activation barrier of a given reaction, $E_a(i)$, is the total electronic energy difference between the transition state (saddle point on a potential energy surface) and the reactants (or product). Reactants, products and transition states were localized on the potential energy surface (PES) using WB97X-D DFT functional. Potential energy surfaces of the reaction pathways relevant to the $N_2$ fixation process can be found in Fig. 4. The reactions of $B_{12}N_{12}$ and $B_{24}N_{24}$ monocyclic rings folding into the corresponding fullborenes can be found in the supplementary materials.

The pre-exponential factors used in calculating the rate constants were calculated using Eyring equation [27],

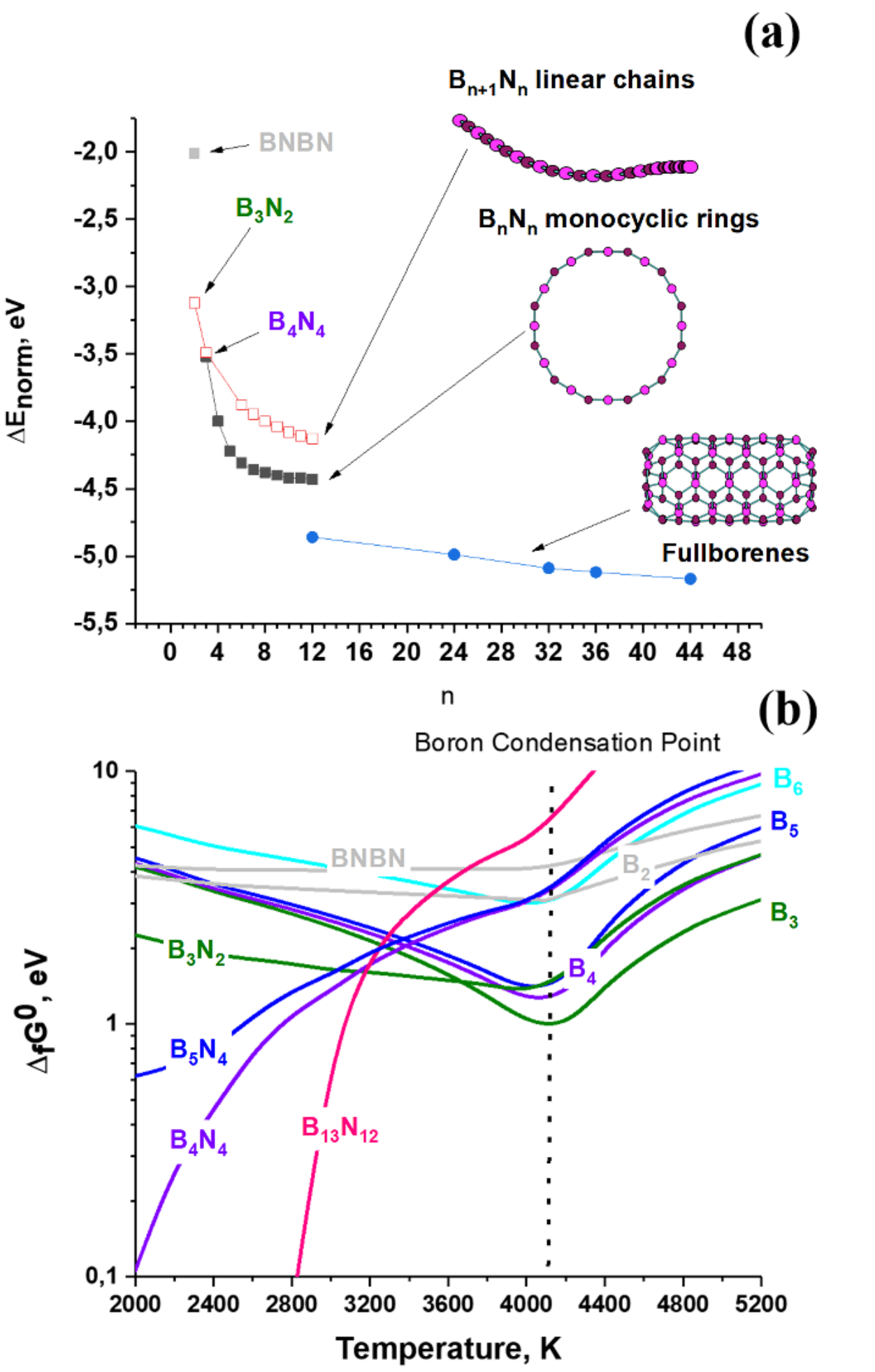

**Figure 2.** Binding energy of $B_mN_n$ clusters ($\Delta E_{norm}$) per reactant (m = n or n+1), dependent on the cluster size and geometry (a). Calculated standard Gibbs free energies of formation (b).

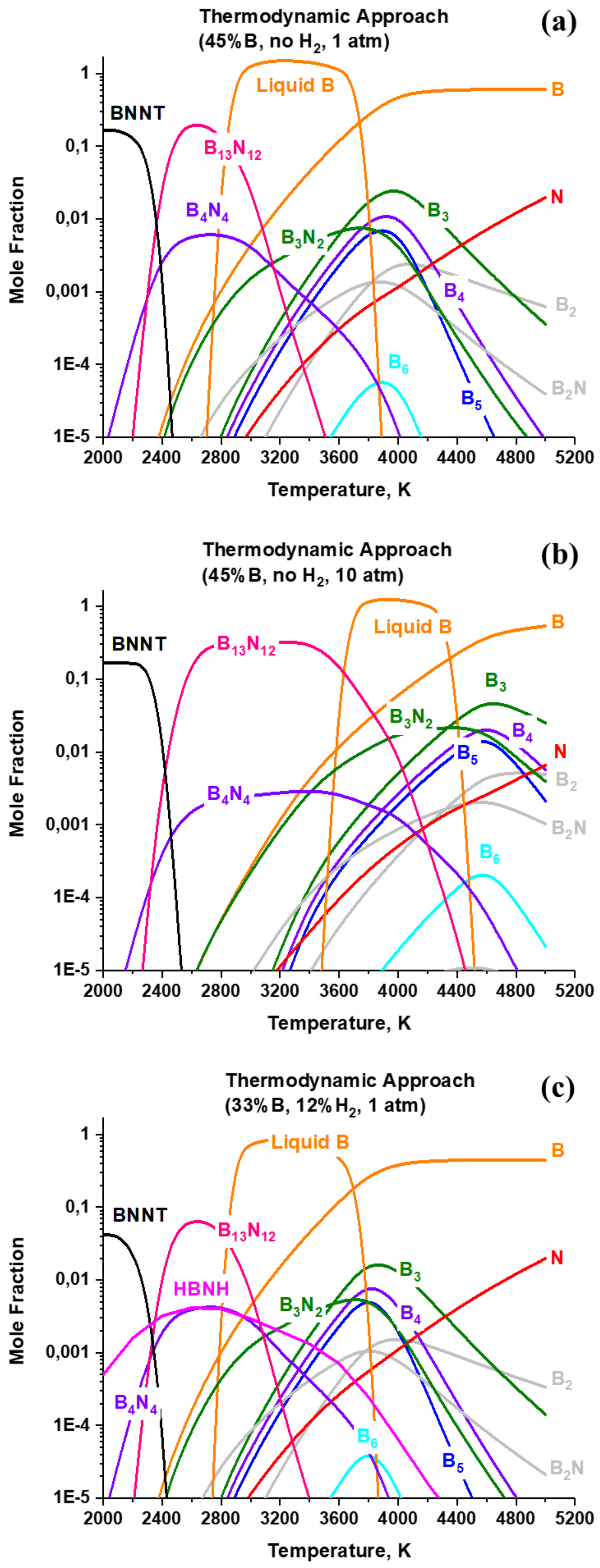

**Figure 3.** Typical equilibrium densities for a variety of molecules for (a) a $B/N_2$ mixture at 1 atm, (b) a $B/N_2$ mixture at 10 atm, and (c) a $B/N_2/H_2$ mixture at 1atm.

$$A`_i = \frac{k_B T}{h} \times \left(\frac{Z_{tot}{}^{TS}}{Z_{tot}{}^{Re}}\right), \qquad (2)$$

where $k_B$, $T$ and $h$ are Boltzmann's constant, Planck's constant, and temperature, respectively. $Z_{tot}{}^{TS}$ and $Z_{tot}{}^{Re}$ are the overall partition functions of the transition state and reactant geometries. $A`_i$ is a function of temperature; it was fitted using the standard form in a temperature range from 2000 to 6000 K:

$$A`_i = A_i\left(\frac{T}{298.15}\right)^n. \qquad (3)$$

The partition functions were calculated using a harmonic oscillator approximation by the GAMESS software package.

A schematic diagram of the calculation methods used in our research is presented in Fig. 1. A DFT approach was used to study the reactions leading to $N_2$ molecular fixation and the process of $B_nN_n$ chains folding into the corresponding fullborene structure, as well as to calculate rate constants for these reactions and standard Gibbs free energies of formation ($\Delta_f G^0$) for the new B-N compounds. The calculated $\Delta_f G^0$ and the rate constants were used as parameters in thermodynamic and kinetic modelling, respectively.

Quantum-classical molecular dynamics (QCMD) was used to investigate reactions involving the merging of $B_3N_3$ rings into longer chains and fullborenes. The self-consistent charge density functional tight binding (SCC-DFTB) approximation was used while performing these simulations. All SCC-DFTB [28] simulations were conducted using the DFTB+ [29] software package. The "matsci-0-3" parameter sets were used for the simulations. These parameter sets are created using semi-relativistic, self-consistent charge Slater-Koster tables [30,31]. Periodic Boundary Conditions and Nose-Hoover thermostat were applied for all calculations. The results of QCMD simulations were visualized using the software Jmol [32].

## 2.1 *Thermodynamic analysis: stability of B-N clusters*

Before investigating the composition of the gas mixture at thermodynamic equilibrium, we first discuss general trends in the stability of B-N monocyclic rings, linear chains, and fullborenes. The geometries of the studied species are depicted in the supplementary materials and in Fig. 2a. We begin by considering the initial reactants which eventually produce these B-N clusters through the nitrogen fixation process:

$$\mathrm{mB} + \frac{n}{2}N_2 \rightarrow B_mN_n. \quad (4)$$

For a variety of B-N monocyclic rings, linear chains, and fullborenes, we performed DFT calculations of the binding energy per reactant, defined as:

$$\Delta E_{norm} = \frac{E(B_mN_n) - mE(B) - \frac{n}{2}E(N_2)}{m + \frac{n}{2}}, \quad (5)$$

where $E(B_mN_n)$, $E(B)$, and $E(N_2)$ are the total electronic energies (calculated using WB97X-D DFT functional) of the product and reactant molecules (corrected for the zero-point vibrational energy of each molecule). A plot depicting the value of $\Delta E_{norm}$ and its dependence on the size of the B-N clusters is shown in Fig. 2a.

The rate of change of $\Delta E_{norm}$ as a function of B-N cluster size is large when the clusters are small. However, the strain of the clusters decreases with elongation, and for long chains ($m>13$), $\Delta E_{norm}$ only weakly depends on the cluster size.

Only chains and rings with m ≤ 13 were considered. We assume that the longer chains will be less stable at high temperatures, given that $\Delta E_{norm}$ weakly depends on chain length for long chains, and that the entropy-related factor in the standard Gibbs free energy of the chains should increase as the length of the chain increases. Carbon chains of comparable size have been observed in Ref. [33–36]. In these studies carbon clusters $C_n$ were generated by laser desorption, after which various isomers of $C_n$ were identified by mass-spectrometry and ion chromatography. This analysis showed that carbon chains with linear structure exist up to $C_{10}^+$. Ref. [37] performed a similar investigation of the equilibrium mixture composition with carbon chains smaller than n=9.

The standard Gibbs free energies of gas phase $B_mN_n$ species depend on $\Delta E_{norm}$ according to the following equation [31]:

$$\Delta_f G^0(B_mN_n) = \left(m + \frac{n}{2}\right)\Delta E_{norm} - k_B T[ln(Z(B_mN_n)) - m * \ln(Z(B)) - \frac{n}{2} * \ln(Z(N_2))] + m * \Delta_f G^0(B(gas)), \quad (6)$$

where Z is the total partition function, $k_B$ is the Boltzmann constant, and $\Delta_f G^0(B(gas))$ is given in the NIST database [18].

The contribution of $\Delta E_{norm}$ into $\Delta_f G^0$ is significantly greater for species with $m \geq 4$. (See Fig. 2a.) As a result, the standard Gibbs free energy of larger species like $B_4N_4$ and $B_5N_4$ monotonically decreases as temperature decreases, as is evident in Fig. 2b. In contrast, for smaller species such as BNBN and $B_3N_2$, the standard Gibbs free energy is a non-monotonic function of temperature and has a minimum near the boron condensation point.

## 2.2 *Thermodynamic modeling: equilibrium mixture composition*

Having calculated the standard Gibbs free energy for each B-N species, we then minimized the total Gibbs free energy to determine the composition of the gas mixture at thermodynamic equilibrium for a variety of different temperatures; for further details see [14]. The results of this thermodynamic modeling are shown in Figure 3.

The densities of long ($m \geq 4$) $B_mN_n$ chains peak in a narrow temperature range, approximately between 2400 and 2800 K at 1 atm, and between 2400 and 3600 at 10 atm.

For the purpose of the thermodynamic modeling, we used $B_{44}N_{44}$ fullborenes to represent BNNTs. Real BNNTs are significantly larger than the $B_{44}N_{44}$ clusters, and are more thermodynamically stable at low temperatures. Thus, BNNTs are more likely final products. However, the $B_{44}N_{44}$ clusters are suitable proxies for larger BNNT molecules in these thermodynamic calculations, as their formation begins at T < 2600 K, which corresponds to the temperature suitable for BNNT formation (Fig. 3).

$B_m$ clusters with $m \leq 5$ have much higher densities than larger $B_m$ clusters for all temperatures. Thus, only these species are considered as sources of boron in our modeling of the $N_2$ fixation process. As shown in the next sections, these boron clusters participate in reactions with $N_2$ to form BNBN, $B_3N_4$, $B_4N_4$ and $B_5N_4$ molecules.

The thermodynamic calculations show that liquid boron is fully consumed by the time the gas has cooled to the boron solidification temperature (~2300 K [18]). (See Fig. 3.) The equilibrium mixture composition at 1 atm and 10 atm is shown in Fig. 3a and 3b. Note that liquid boron disappears at temperatures below 2800K at 1 atm and below 3600 K at 10 atm. This demonstrates that higher pressures promote liquid boron consumption.

The addition of $H_2$ into the B/$N_2$ mixture leads to the formation of a significant amount of HBNH (iminoborane) molecules, as shown in Fig. 3c. Note that similar to the B-N rings and chains, HBNH density increases with decreasing temperature until BNNT growth begins. Correspondingly, this molecule can play a key role in the synthesis of BNNTs in the presence of hydrogen. (See discussion section.)

## 2.3 *Reaction pathways for $N_2$ fixation via interaction with $B_2$, $B_3$, $B_4$, and $B_5$ clusters*

The $B_m$ clusters with m≤5 have the highest densities after boron condensation, as discussed in the previous section. Here, we present a detailed chemical mechanism for $N_2$ fixation via reactions with these clusters and describe the relevant energies and geometries along the reaction pathways.

Calculated reaction paths of nitrogen fixation resulting in the production of BNBN, $B_3N_2$, $B_4N_4$ and $B_5N_4$ are shown in Fig. 4. It requires several steps to achieve the most stable geometry, which corresponds to a global minimum on the potential energy surface. The highest barriers to dissociation for BNBN, $B_3N_2$, $B_4N_4$ and $B_5N_4$ are significantly larger for the larger molecules (2.5, 4.8, 5.8 and 6.0 eV, respectively).

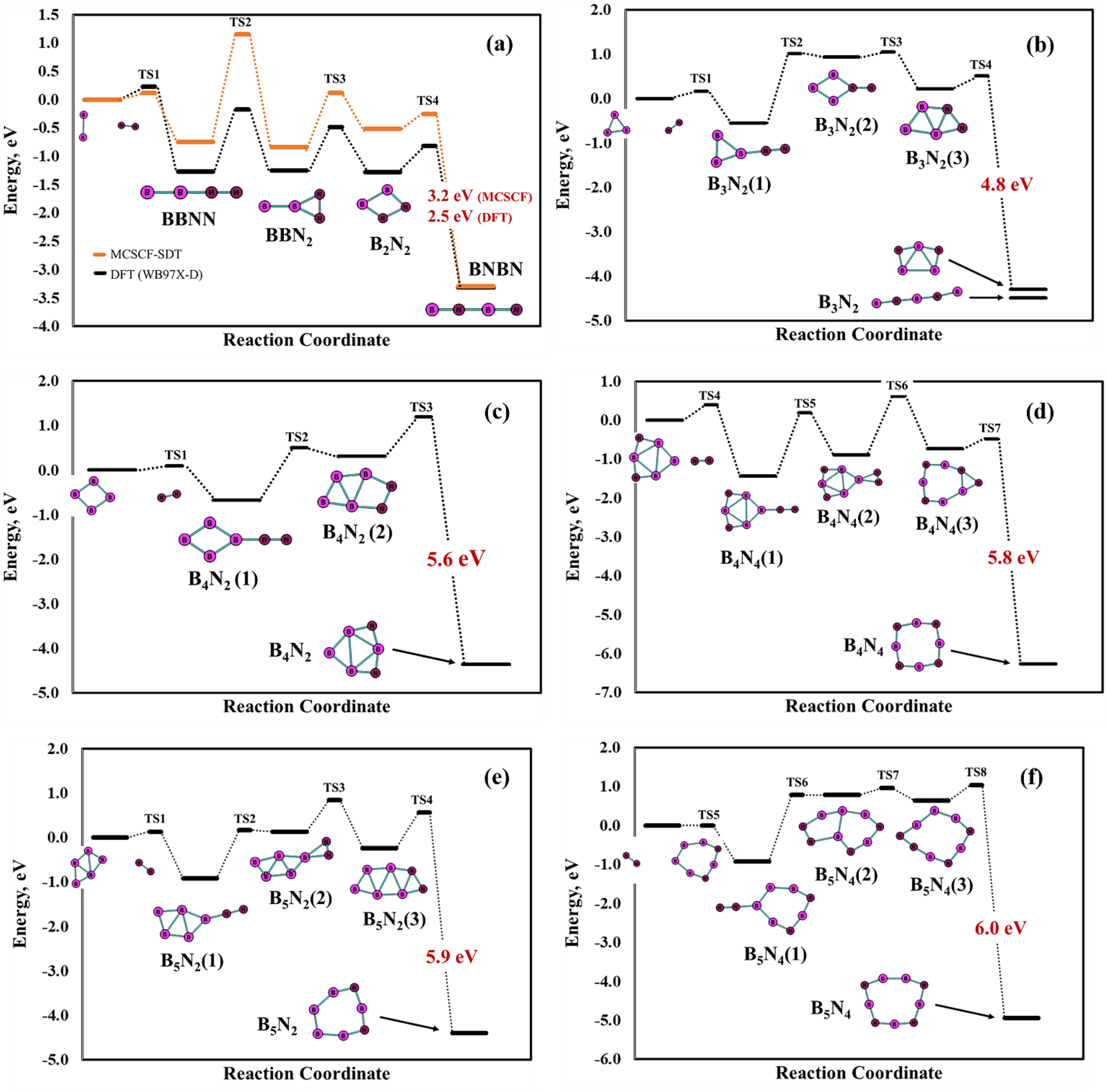

**Figure 4.** Calculated reaction paths of $N_2$ molecule fixation in the reactions with $B_2$ (a), $B_3$ (b), $B_4$ ((c) and (d)), and $B_5$ ((e) and (f)).

The reaction path for the transformation of $B_2$ and $N_2$ dimers into BNBN is depicted in Figure 4a. Four transition states (TSs) have been localized (TS1, TS2, TS3, TS4) using both DFT and MCSCF methods. The first reaction step is the formation of a linear BBNN molecule, which proceeds through a fairly low energy barrier (~0.23 eV). It should be noted that the BBNN molecule structure is located in a deep local minimum on the potential energy surface (Fig. 4a). This explains experimental data presented in [32], where BBNN molecule formation has been observed in addition to the formation of the more stable BNBN molecule. The second step through TS2 is the highest barrier along the reaction pathway in the forward direction, and results in the formation of a planar $BBN_2$ molecule. In the third step the $BBN_2$ molecule is transformed into a planar $B_2N_2$ molecule through TS3. The $B_2N_2$ molecule is the most unstable isomer and it is transformed into the linear BNBN molecule through TS4. Note that both DFT and MCSCF approaches reproduce qualitatively the same reaction path for BNBN molecule formation. For reactions with larger clusters shown in Fig. 4b-4f, only a DFT approach was used.

This modelling demonstrates that reactions between small $B_m$ clusters $N_2$ dimers result in nitrogen fixation and B-N molecule generation without the presence of atomic nitrogen. Such mechanisms require less energy than the atomization of $N_2$ dimers; while the $N_2$ bond energy is 9.8 eV, the highest barrier towards BNBN molecule formation from the proposed series of reactions described above is only 1.10 eV.

Calculated rate constants (see Eq. 1) for the reaction steps presented in Fig. 4 can be found in the supplementary materials. These rate constants are used in the kinetic modeling described in the next section.

### 2.4 Kinetic modeling of $N_2$ fixation

In this section we consider kinetic modeling of the formation of BNBN, $B_3N_2$, $B_4N_4$ and $B_5N_4$ species through reactions between $N_2$ molecules and $B_2$, $B_3$, $B_4$ and $B_5$ clusters, respectively. Based on the chemical reaction pathways established in the previous section, reaction rate constants were determined and the kinetic equations describing the time evolution of species' densities were numerically solved. The systems of differential equations describing BNBN, $B_3N_2$, $B_4N_4$, and $B_5N_4$ formation are presented below.

**BNBN generation:**

$$\frac{dC_{BBNN}(t)}{dt} = \overrightarrow{k1}C_{B2}^{eq}(t)C_{N2}(t) - \left(\overleftarrow{k1} + \overrightarrow{k2}\right)C_{BBNN}(t) + \overleftarrow{k2}C_{BBN2}(t)$$

$$\frac{dC_{BBN2}(t)}{dt} = \overrightarrow{k2}C_{BBNN}(t) - \left(\overleftarrow{k2} + \overrightarrow{k3}\right)C_{BBN2}(t) + \overleftarrow{k3}C_{B2N2}(t)$$

$$\frac{dC_{B2N2}(t)}{dt} = \overrightarrow{k3}C_{BBN2}(t) - \left(\overleftarrow{k3} + \overrightarrow{k4}\right)C_{B2N2}(t) + \overleftarrow{k4}C_{BNBN}(t)$$

$$\frac{dC_{BNBN}(t)}{dt} = \overrightarrow{k4}C_{B2N2}(t) - \overleftarrow{k4}C_{BNBN}(t) \qquad (7a)$$

**$B_3N_2$ generation**:

$$\frac{dC_{B3N2(1)}(t)}{dt} = \overrightarrow{k1}C_{B3}^{eq}(t)C_{N2}(t) - \left(\overleftarrow{k1} + \overrightarrow{k2}\right)C_{B3N2(1)}(t) + \overleftarrow{k2}C_{B3N2(2)}(t)$$

$$\frac{dC_{B3N2(2)}(t)}{dt} = \overrightarrow{k2}C_{B3N2(1)}(t) - \left(\overleftarrow{k2} + \overrightarrow{k3}\right)C_{B3N2(2)}(t) + \overleftarrow{k3}C_{B3N2(3)}(t)$$

$$\frac{dC_{B3N2(3)}(t)}{dt} = \overrightarrow{k3}C_{B3N2(2)}(t) - \left(\overleftarrow{k3} + \overrightarrow{k4}\right)C_{B3N2(3)}(t) + \overleftarrow{k4}C_{B3N2}(t)$$

$$\frac{dC_{B3N2}(t)}{dt} = \overrightarrow{k4}C_{B3N2(3)}(t) - \overleftarrow{k4}C_{B3N2}(t) \qquad (7b)$$

**$B_4N_4$ generation:**

$$\frac{dC_{B4N2(1)}(t)}{dt} = \overrightarrow{k1}C_{B4}^{eq}(t)C_{N2}(t) - \left(\overleftarrow{k1} + \overrightarrow{k2}\right)C_{B4N2(1)}(t) + \overleftarrow{k2}C_{B4N2(2)}(t)$$

$$\frac{dC_{B4N2(2)}(t)}{dt} = \overrightarrow{k2}C_{B4N2(1)}(t) - \left(\overleftarrow{k2} + \overrightarrow{k3}\right)C_{B4N2(2)}(t) + \overleftarrow{k3}C_{B4N2(3)}(t)$$

$$\frac{dC_{B4N2}(t)}{dt} = \overrightarrow{k3}C_{B4N2(2)}(t) - \left(\overleftarrow{k3} + \overrightarrow{k4}C_{N2}(t)\right)C_{B4N2}(t) + \overleftarrow{k4}C_{B4N2}(t)$$

$$\frac{dC_{B4N4(1)}(t)}{dt} = \overrightarrow{k4}C_{N2}(t)C_{B4N2}(t) - \left(\overleftarrow{k4} + \overrightarrow{k5}\right)C_{B4N4(1)}(t) + \overleftarrow{k5}C_{B4N4(2)}(t)$$

$$\frac{dC_{B4N4(2)}(t)}{dt} = \overrightarrow{k5}C_{B4N4(1)}(t) - \left(\overleftarrow{k5} + \overrightarrow{k6}\right)C_{B4N4(2)}(t) + \overleftarrow{k6}C_{B4N4(3)}(t)$$

$$\frac{dC_{B4N4(3)}(t)}{dt} = \overrightarrow{k6}C_{B4N4(2)}(t) - \left(\overleftarrow{k6} + \overrightarrow{k7}\right)C_{B4N4(3)}(t) + \overleftarrow{k7}C_{B4N4}(t)$$

$$\frac{dC_{B4N4}(t)}{dt} = \overrightarrow{k7}C_{B4N4(3)}(t) - \overleftarrow{k7}C_{B4N4}(t) \qquad (7c)$$

**$B_5N_4$ generation:**

$$\frac{dC_{B5N2(1)}(t)}{dt} = \overrightarrow{k1}C_{B5}^{eq}(t)C_{N2}(t) - \left(\overleftarrow{k1} + \overrightarrow{k2}\right)C_{B5N2(1)}(t) + \overleftarrow{k2}C_{B5N2(2)}(t)$$

$$\frac{dC_{B5N2(2)}(t)}{dt} = \overrightarrow{k2}C_{B5N2(1)}(t) - \left(\overleftarrow{k2} + \overrightarrow{k3}\right)C_{B5N2(2)}(t) + \overleftarrow{k3}C_{B5N2(3)}(t)$$

$$\frac{dC_{B5N2(3)}(t)}{dt} = \overrightarrow{k3}C_{B5N2(2)}(t) - \left(\overleftarrow{k3} + \overrightarrow{k4}\right)C_{B5N2(3)}(t) + \overleftarrow{k4}C_{B5N2}(t)$$

$$\frac{dC_{B5N2}(t)}{dt} = \overrightarrow{k4}C_{B5N2(3)}(t) - \left(\overleftarrow{k4} + \overrightarrow{k5}C_{N2}(t)\right)C_{B5N2}(t) + \overleftarrow{k5}C_{B4N4(1)}(t)$$

$$\frac{dC_{B5N4(1)}(t)}{dt} = \overrightarrow{k5}C_{N2}(t)C_{B5N2}(t) - \left(\overleftarrow{k5} + \overrightarrow{k6}\right)C_{B5N4(1)}(t) + \overleftarrow{k6}C_{B5N4(2)}(t)$$

$$\frac{dC_{B5N4(2)}(t)}{dt} = \overrightarrow{k6}C_{B5N4(1)}(t) - \left(\overleftarrow{k6} + \overrightarrow{k7}\right)C_{B5N4(2)}(t) + \overleftarrow{k7}C_{B5N4(3)}(t)$$

$$\frac{dC_{B5N4(3)}(t)}{dt} = \overrightarrow{k7}C_{B5N4(2)}(t) - \left(\overleftarrow{k7} + \overrightarrow{k8}\right)C_{B5N4(3)}(t) + \overleftarrow{k8}C_{B5N4}(t)$$

$$\frac{dC_{B5N4}(t)}{dt} = \overrightarrow{k8}C_{B5N4(3)}(t) - \overleftarrow{k8}C_{B5N4}(t) \qquad (7d)$$

In these systems of equations, the rate constants of forward and reverse reactions ($\overrightarrow{ki}$ and $\overleftarrow{ki}$) are related to the $TS_i$ transition states in Fig. 4. The rate constants depend on temperature, which in turn is assumed to be linearly decreasing with time via a constant gas cooling rate ($\dot{T}_0$):

$$T(t) = T_0 - \dot{T}_0 \times t. \qquad (8)$$

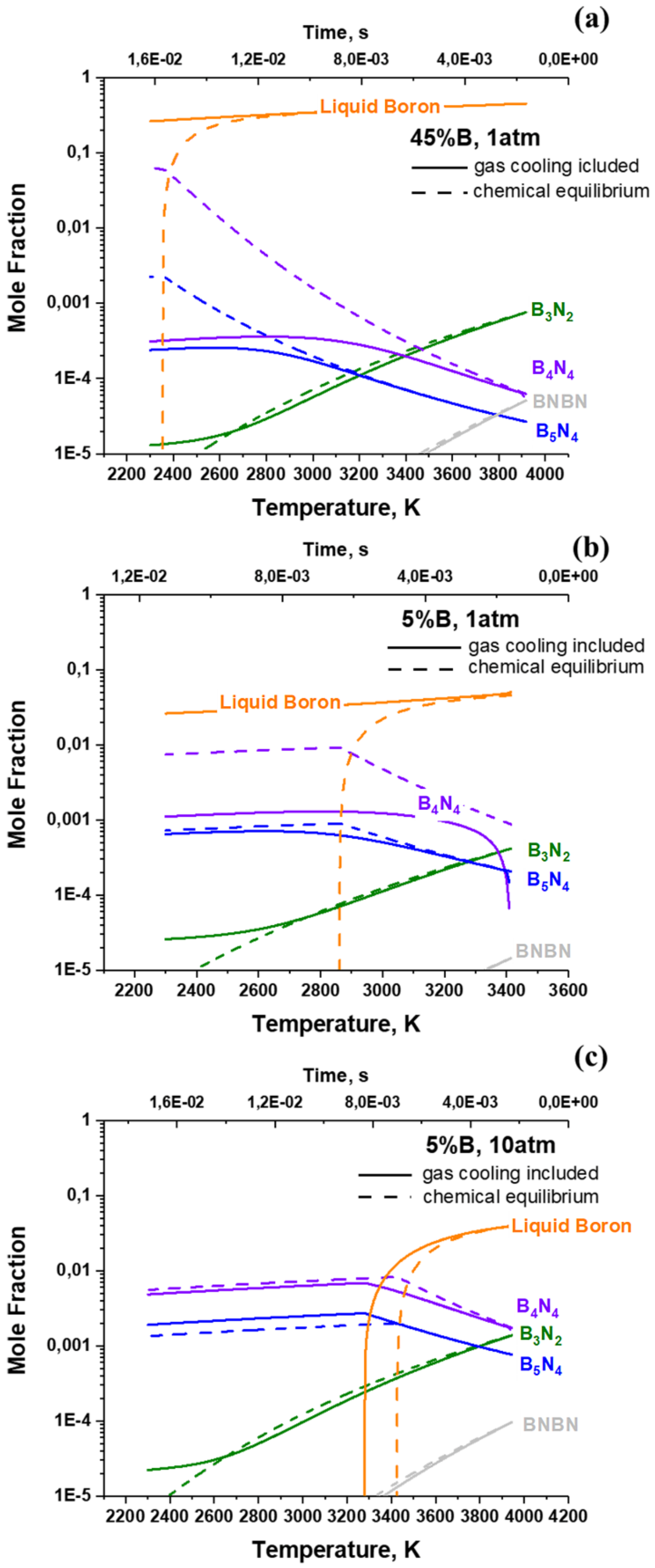

**Figure 5.** $N_2$ fixation process during a gas cooling under kinetic control ($\dot{T}_0$=$10^5$ K/s) at different conditions: (a) 45%B and 1 atm, (b) 5%B and 1 atm, (c) 5%B and 10 atm.

Thus, both the rate constants and equilibrium densities of $B_m$ clusters depend on temperature, which in turn depends on a gas cooling rate. Essentially, both rate constants as well as densities of $B_m$ clusters depend on time.

The equilibrium densities of $B_m$ clusters were obtained using the following equation:

$$C^{eq}_{Bm}(t) = C^{eq}_B(t)^m \times K_f(B_m), \tag{9}$$

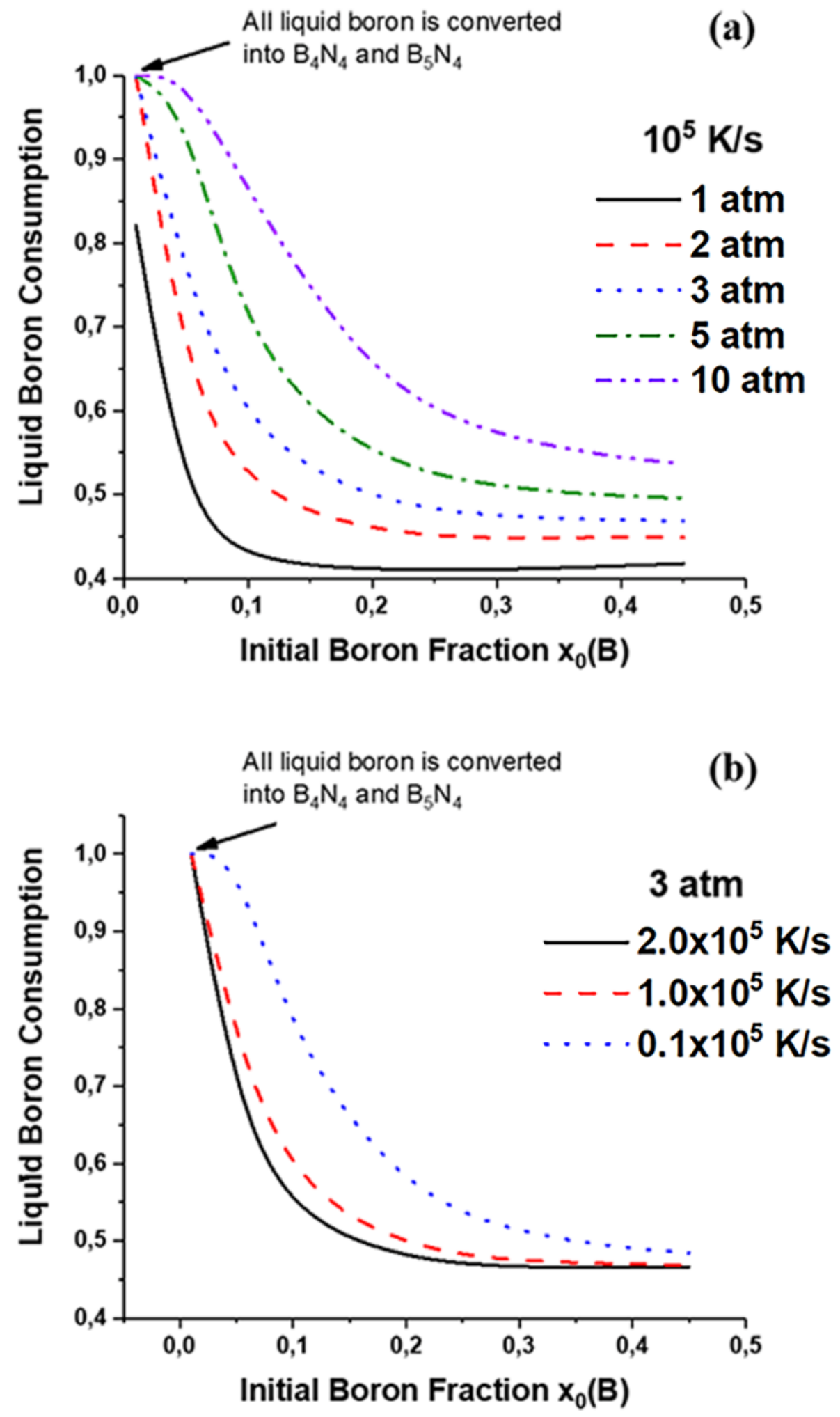

**Figure 6.** Liquid boron consumption (LBC) vs initial boron fraction at (a) different pressures and (b) gas cooling rates.

where $K_f(B_m) = \exp\left(-\frac{\Delta G(B_m)}{RT}\right)$ is the calculated equilibrium constant of $B_m$ formation from boron atoms in gas and $C^{eq}_B$ is the equilibrium density of boron atoms in gas phase. The latter was calculated according to the Clausius-Clapeyron relation:

$$C^{eq}_B(t) \approx \frac{P}{RT(t)} \times \exp\left(-\frac{\Delta_f G(B,T(t))}{RT(t)}\right), \tag{10}$$

where $\Delta_f G(B,T(t)) = \Delta_f G^0(B,T(t)) + RT(t)ln(\frac{P}{P_0})$ and $\Delta_f G^0(B,T(t))$ is the standard Gibbs free energy of formation for atomic boron at 1 atm.

The densities of BNBN, $B_3N_2$, $B_4N_4$ and $B_5N_4$ at chemical equilibrium ($C^{eq}$) were calculated using the system of equations 7a-7d, taking into account that $\frac{dC(t)}{dt} = 0$ for all species:

$$C^{eq}_{BNBN}(T(t)) = \frac{\sum_{i=1}^{4}\vec{k}_i}{\sum_{i=1}^{4}\overleftarrow{k}_i} C^{eq}_{B2}(T(t)) C^{eq}_{N2}(T(t)) = \frac{\sum_{i=1}^{4}\vec{k}_i}{\sum_{i=1}^{4}\overleftarrow{k}_i} C^{eq}_{B2}(T(t))(1 - x_0(B))\frac{P}{RT}. \tag{11a}$$

$$C^{eq}_{B3N2}(T(t)) = \frac{\sum_{i=1}^{4}\vec{k}_i}{\sum_{i=1}^{4}\overleftarrow{k}_i} C^{eq}_{B3}(T(t)) C^{eq}_{N2}(T(t)) = \frac{\sum_{i=1}^{4}\vec{k}_i}{\sum_{i=1}^{4}\overleftarrow{k}_i} C^{eq}_{B3}(T(t))(1 - x_0(B)) \frac{P}{RT}. \quad (11b)$$

$$C^{eq}_{B4N4}(T(t)) = \frac{\sum_{i=1}^{7}\vec{k}_i}{\sum_{i=1}^{7}\overleftarrow{k}_i} C^{eq}_{B4}(T(t)) C^{eq}_{N2}(T(t))^2 = \frac{\sum_{i=1}^{7}\vec{k}_i}{\sum_{i=1}^{7}\overleftarrow{k}_i} C^{eq}_{B4}(T(t))\left(1 - x_0(B)\right)^2 (\frac{P}{RT})^2. \quad (11c)$$

$$C^{eq}_{B5N4}(T(t)) = \frac{\sum_{i=1}^{8}\vec{k}_i}{\sum_{i=1}^{8}\overleftarrow{k}_i} C^{eq}_{B5}(T(t)) C^{eq}_{N2}(T(t))^2 = \frac{\sum_{i=1}^{8}\vec{k}_i}{\sum_{i=1}^{8}\overleftarrow{k}_i} C^{eq}_{B5}(T(t))\left(1 - x_0(B)\right)^2 (\frac{P}{RT})^2. \quad (11d)$$

The condensation point ($T_B$) depends on boron partial pressure ($x_0(B)P$). We define $T_B$ as a temperature at which $\Delta_f G(B, T(t))$=0.

The results of the kinetic modeling are presented in Fig. 5 for the temperature range between the boron solidification (2300 K) and condensation points ($T_B$). The fraction of liquid boron ($x_{LB}$) is a fraction of boron atoms in liquid state, which was defined as

$$x_{LB} = \left(x_0(B)\frac{PV}{RT_B} - 4C_{B4N4}(t) - 5C_{B5N4}(t)\right)\frac{RT(t)}{PV}. \quad (12)$$

It is evident from Fig. 5 that the fractions of $B_4N_4$ and $B_5N_4$ species deviate significantly from the chemical equilibrium values at T < $T_B$ for atmospheric and higher pressures. As a result, kinetic simulations predict that liquid boron is not fully consumed before boron solidification at 1 atm pressure (Fig. 5a and 5b), assuming that the typical gas cooling rate during BNNT synthesis is $10^5$ K/s [8]. This differs from the complete liquid boron consumption predicted by earlier thermodynamic calculations in Section 2.2. According to equations 11c and 11d, high pressure (10 atm) and a low initial boron fraction ($x_0(B)$ =5%) is required in order to significantly enhance $B_4N_4$ and $B_5N_4$ production via liquid boron consumption (Fig. 5c). Liquid boron consumption (LBC) is an important issue to consider for the production of high quality BNNTs. (See discussion section). Here, the LBC at the boron solidification point (2300 K) has been estimated using the following expression:

$$LBC = \frac{x_0(B) - x_{LB}(2300K)}{x_0(B)}. \quad (13)$$

As shown in Fig. 6, LBC increases with higher pressures and slower gas cooling rates. Total consumption of all liquid boron is possible only with a low initial boron fraction and at pressures higher than 1 atm (Fig. 6a). Indeed, even with an initial boron fraction of $x_0(B)$=0.01, not all liquid boron will be consumed if the pressure is only 1 atm. This insight agrees with experimental observations of boron droplets encapsulated in BN shell structures at 1 atm [7]. (See discussion.)

The effect of the gas cooling rate on LBC is effectively demonstrated in Fig. 6b. It is shown that the gas cooling should be 10 times slower (~$10^4$ K/s) than the typical cooling rate in order to significantly increase LBC.

## 2.5 QCMD simulation of small chains aggregation

We used DFTB+ modeling to investigate the interaction of small $B_3N_3$ rings with each other. We showed that planar collisions of these $B_3N_3$ molecules lead to the formation of one big monocyclic $B_{12}N_{12}$ ring. This result supports our thermodynamic modeling, which indicated that longer B-N chains, both cyclical and linear, are more stable than shorter B-N chains at T < 3000 K. We suggest that $B_4N_4$ and $B_5N_4$ clusters will aggregate into larger chains through mechanisms similar to the one explored in the case of $B_3N_3$. Nonplanar collisions of these rings lead to the formation of $B_{12}N_{12}$ and $B_{24}N_{24}$ fullborenes in the shape of nanocages [38]. Thus, the smaller B-N rings tend to combine very quickly (on a nanosecond timescale) into longer chains and fullborenes. This confirms our assumption that the rate-limiting step of BNNT precursor formation is the $N_2$ fixation process resulting in formation of small $B_mN_n$ chains ($B_4N_4$ and $B_5N_4$). Once these small chains form, the production of longer chains and subsequent synthesis of fullborenes and BNNTs proceeds rapidly during a gas cooling.

## 2.6 Rings folding into fullborenes

A detailed mechanism of BNNT formation from the B-N precursors is beyond the scope of this research. However, the reaction paths for the transformation of $B_{12}N_{12}$ and $B_{24}N_{24}$ chains (monocyclic rings) into the corresponding fullborene structures have been calculated (Fig. 7). The rate constants corresponding to these transformations can be found in supplementary materials. The $B_{12}N_{12}$ fullborene formation is shown in Fig. 7a. Five steps have been found on this reaction path. The first three reaction steps are the transformation of the monocyclic ring into a planar graphitic sheet, and the following two steps involve the folding of the B-N plane into a 3D fullborene via TS4 and TS5.

The formation of a $B_{24}N_{24}$ fullborene proceeds through 16 transition states (Fig. 7b). The geometries of the 15 intermediate structures are shown in Fig. 7c. The first 12 steps describe the formation of a planar graphitic structure consisting of $B_3N_3$ hexagonal rings (see $B_{24}N_{24}$ (12) in Fig. 7b.) The subsequent steps describe the transformation of the planar structure into a "cup" shaped structure (see $B_{24}N_{24}$ (15) in Fig. 7b.) and eventually into the 3D fullborene. All barriers of the forward reactions are less than 3.1 eV, and the geometry of the $B_{24}N_{24}$ fullborene corresponds to a very deep well on the potential energy surface (10.5 eV), which is higher than the bond energy of $N_2$.

It should be noted that our reaction paths for fullborene formation bear a resemblance to the mechanism of fullerene formation proposed in [33–35]. In that study, various $C_n$ cations were generated by laser desorption and identified based on mass-spectrometry, ion chromatography data, and quantum chemistry calculations [34–36]. From this analysis, it was determined that the growth of fullerenes follows the following reaction steps: chains → monocyclic rings → polycyclic rings → fullerenes [35]. According to our simulations, small chains form longer monocyclic rings via collisions with each other (Section 2.5), and the monocyclic rings fold into fullborenes via a series of transformations very similar to those described above (Fig. 7): monocyclic ring → bicyclic ring → polycyclic rings → graphitic sheet → “cup” shaped structure → fullborene.

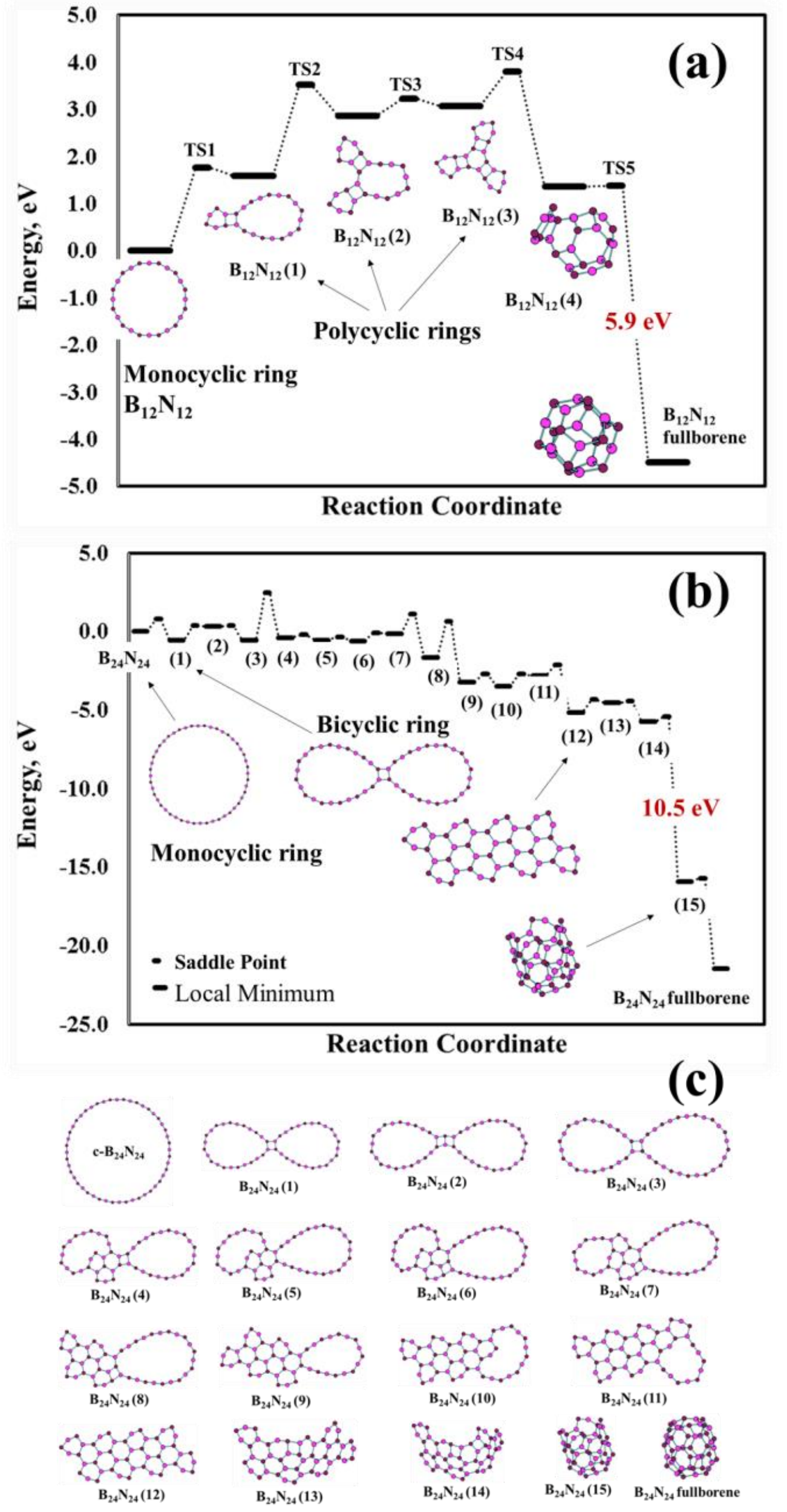

**Figure 7.** Calculated reaction path (potential energy surface) of transformation of (a) $B_{12}N_{12}$ and (b) $B_{24}N_{24}$ monocyclic rings into the corresponding fullborenes. (c) The reactant, product and intermediate structures of the reaction path of transformation of $B_{24}N_{24}$ monocyclic ring into the corresponding fullborene

The graphitic and “cup” shaped structures described by our modelling of fullborene growth were not observed in the fullerene growth experiment [35]. This discrepancy can be explained by the low energy barriers along the reaction path near these structures. Indeed, according to our DFT calculations, these structures ($B_{24}N_{24}$ (12) and $B_{24}N_{24}$ (15) in Fig. 7b.) are transformed into fullborenes through TS13-TS15, and the barriers corresponding to these transitions are quite low, while the barriers to backwards reactions from the fullborene structures are quite large. Thus, the graphitic and "cup" shaped isomers are most likely short-lived intermediates on the reaction path leading to fullborene (or fullerene) formation and therefore difficult to detect experimentally.

It is interesting to note that the smallest $C_n$ fullerene which observed in [33,35] has n=30. In addition, the vast majority of $C_n$ clusters with n>50 have folded into fullerenes [35]. This phenomenon can be explained by comparing the reaction barriers in the decomposition of the two fullborenes examined in our modelling. The highest reaction barrier in the decomposition of $B_{24}N_{24}$ is much higher than that of $B_{12}N_{12}$ (5.9 vs. 10.5 eV in Fig. 7a and 7b). Thus, the stability of fullborenes significantly increases with cluster size. Similarly, smaller fullerenes should be less stable than bigger ones, which explains the fact that $C_n$ fullerenes with n<30 were not observed in [33,35].

## 3 Discussion

As we discussed in Section 2.6 the fullborene formation during high temperature BNNT synthesis should proceed through the formation of the following structures: small B-N chains → monocyclic rings → bicyclic rings → polycyclic rings → graphitic sheet → “cup” shaped structure → fullborene. Our research aims to study in detail the synthesis of the first structure (small B-N chains) through $N_2$ fixation.

Here it has been shown that $N_2$ molecules can react with $B_m$ clusters to produce small $B_mN_n$ chains, and the main reaction products are $B_4N_4$ and $B_5N_4$ chains (Section 2.4). The source of these $B_m$ clusters is liquid boron, and thus the consumption of these $B_m$ clusters by the nitrogen fixation process results in the consumption of liquid boron. Under the (not valid) assumption that thermodynamic equilibrium is achieved throughout gas cooling process, all liquid boron is consumed. (See Fig. 3.) Removing the assumption of thermodynamic equilibrium, we show that consumption of all liquid boron is only achieved under the conditions of high pressure and low initial boron fraction. Correspondingly, the formation of small $B_mN_n$ chains occurs in a non-equilibrium (kinetic) regime and depends on the gas cooling rate and pressure. The effect of pressure was reported by the Zettl group [7], where it was concluded that boron is not completely consumed by reactions with $N_2$ at 1 atm pressure. As a result, shell structures of BN around boron droplets (“nanococoons”) are observed. At pressures above 2 atm the prevalence of these nanococoons begins to decrease and at pressures above 3 atm nanococoon formation is significantly suppressed, assisting in high purity BNNT synthesis. The observed boron consumption trend is consistent with the kinetic modeling results shown in Fig. 5 and 6, where it can be seen that the liquid boron consumption is hindered at 1 atm pressure. As a result, not all boron is converted into $B_4N_4$ and $B_5N_4$ chains. Higher pressures enhance $B_4N_4$ and $B_5N_4$ production and liquid boron can be fully consumed even at 2 atm (Fig. 6a).

It should be noted that the complicated process of boron droplet nucleation prior to the agglomeration phase is not considered here. Only boron clusters $B_m$ with m≤6 and liquid boron were considered in this study. The clusters were shown to contribute to nitrogen fixation, while the liquid boron did not [19]; contributions from larger $B_m$ clusters were not studied. However, they may also play a role in nitrogen fixation, either directly or indirectly. The cluster size at which these contributions to nitrogen fixation are no longer important has not been determined, though previous studies have established that clusters as large as $B_{13}$ are capable of reacting with $N_2$ [9].

During the nucleation stage, the small $B_m$ clusters are in equilibrium with the bigger boron clusters, not with liquid state as suggested in our current kinetic model. Because the larger boron clusters are not accounted for in our model, the densities of smaller boron clusters may be underestimated near the nucleation point in our calculations. Considering that the nucleation burst typically occurs at temperatures ~300K below the nucleation point [39], we believe that the kinetic model underestimates the rate of $B_4N_4$ generation near the boron boiling point (Fig. 5b) since the nucleation burst is not taken into account.

The study of the growth process of BNNTs is beyond the scope of our research, but some steps were taken in this direction. Namely, we showed using DFTB+ simulations (Section 2.5) that the small B-N chains form longer B-N rings. The reaction paths leading to the folding of B-N rings into fullborenes is shown in in Section 2.6. We propose that the formation of fullerenes and fullborenes occurs through a series of similar structural transformations.

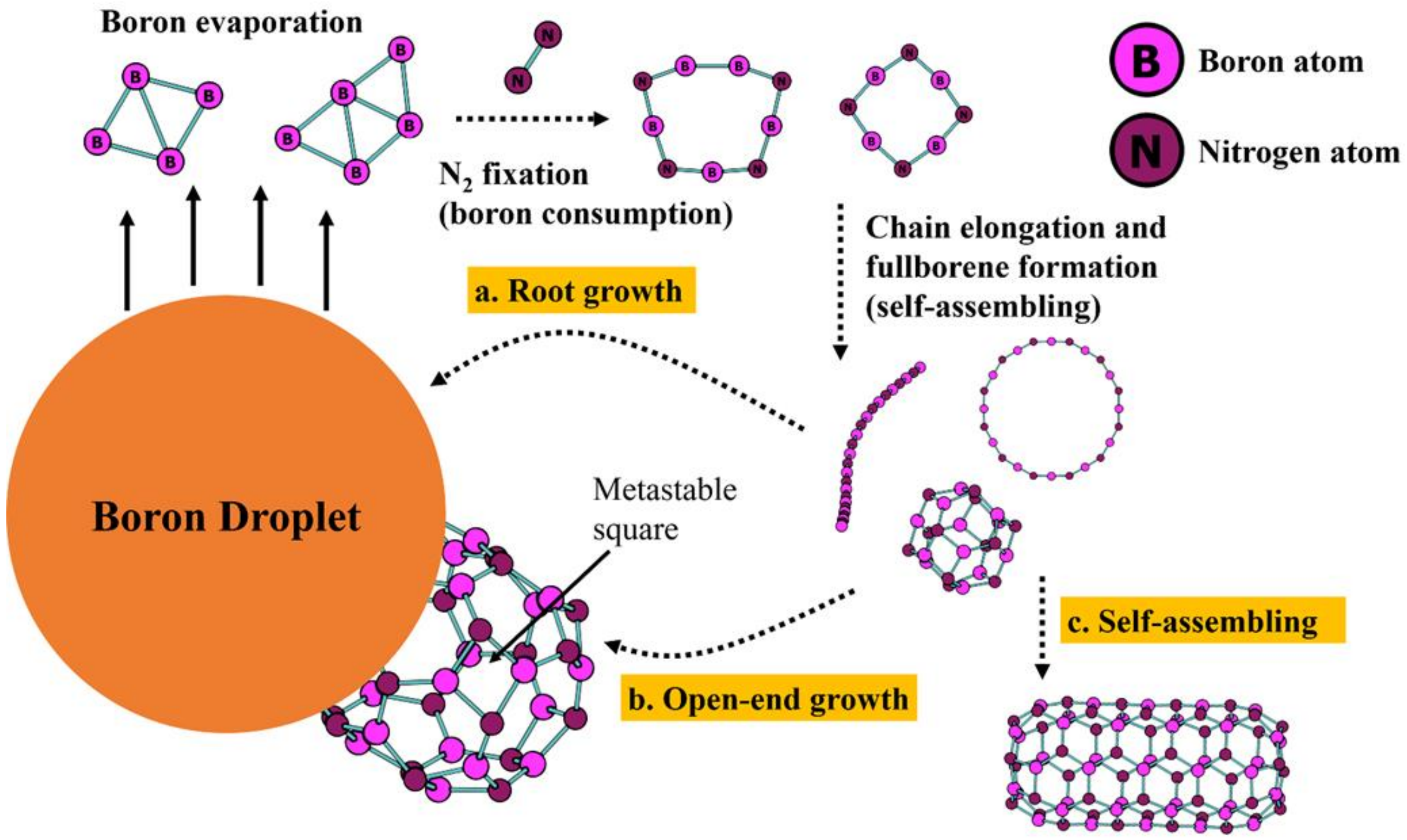

**Figure 8.** Mechanisms of BNNTs growth: (a) root growth, (b) open-end growth and (c) self-assembling.

Despite this structural similarity, fullborenes should be less stable than fullerenes. Blasé et al. [40] showed that fullborene caps consist of metastable squares due to the greater stability of B-N bonds compared to B-B and N-N bonds. In other words, stabilized fullborenes and BNNTs are capped by metastable sites (squares), and can therefore attach to each other by these sites to form elongated structures. This process was reported to occur for the $B_{12}N_{12}$ fullborene [41]. This indicates that fullborenes are less stable than BNNTs, because the longer structure of the nanotubes consists mainly of stable hexagonal rings, so the effect of the metastable sites on the stability of whole tube is insignificant.

Our thermodynamic modelling (Section 2.2) predicts that fullborene formation begins at temperatures less than 2600K, about the same temperature at which BNNT growth begins [9]. Thus, if fullborenes are generated in the gas mixture they will collide and form more stable elongated structures, eventually resulting in the growth of BNNTs. This process resembles the self-assembling mechanism proposed in [17,42]. (Or see Fig. 8c.) These studies have reported that BN dimers, BNH, HBNH, and $B_3N_3H_3$ molecules can self-assemble into BN nanostructures (cages, flakes, and tubes) without the presence of boron droplets. It was also observed that BN chains are the first to appear in the B-N reaction mixture. This is consistent with our result that $B_mN_n$ chains are precursors to the growth of BNNTs.

Note, a way to synthesize the BN chains had not been experimentally discovered despite the fact that potential applications of BN chains were theoretically studied [43]. Moreover, the very fact that BN chains exist had not been discussed until they were experimentally observed above hexagonal boron nitride sheets [44]. According to our data, synthesis of BN chains is a difficult process since the chains are stable only within a narrow temperature range (Section 2.2).

Another BNNT growth mechanism previously explored in the literature [4,8] is the root growth mechanism. In this process, $B_mN_n$ chains or fullborenes are dissolved in boron droplets, and BNNT growth proceeds outwards from the droplet surface as shown in Fig. 8a. The same setup, wherein the BNNT is attached to a boron droplet at one end, also supports another mechanism for BNNT growth. In this process, $B_mN_n$ chains and fullborenes can react with the metastable site at the tube-end, resulting in tube elongation. This resembles the open-end growth mechanism described in [16]. (See Fig. 8b.)

Previously, hydrogen was reported to have a catalytic role in BNNT synthesis [8]. Here, we considered the effect of hydrogen using a thermodynamic approach, and showed that iminoborane (HBNH) molecules can be produced in large quantities. (See Fig. 3c.) It should be noted that HBNH was detected in a diborane/ammonia arc discharge [45] and in reactions between atomic boron and $NH_3$ [46]. The mechanisms of HBNH formation and HBNH incorporation into BNNT molecules were previously studied using quantum chemistry methods [47]. It was shown that HBNH molecules are a possible precursor in the production of borazine [42] as well as BNNTs [42,47]. In addition, as shown in Section 2.6, the graphitic and "cup" shaped structures are not stable. The hydrogen atoms will close the vacancies on the terminal B and N atoms, stabilizing these structures.

## 4 Conclusion

We have proposed a chemical reaction pathway for BNNT growth and precursor formation. We showed that the critical (rate-determining) step in BNNT synthesis is $N_2$ fixation, which primarily occurs during the reaction of $N_2$ with small $B_4$ and $B_5$ clusters, causing the formation of $B_4N_4$ and $B_5N_4$ chains. Subsequent chain elongation occurs via collisions of $B_4N_4$ and $B_5N_4$ with each other, and these larger chains are in turn precursors to fullborene and BNNT growth. The process described here stands in contrast with the common assumption that small molecules such as BN, $B_2N$, and $BN_2$ are the main precursors to BNNT formation.

We also showed that high pressures and slow gas cooling rates enhance $N_2$ fixation and correspondingly enhance liquid boron consumption, creating good conditions for high quality BNNT synthesis. The addition of hydrogen into the reaction mixture can also enhance $N_2$ fixation. It was shown that HBNH molecules are present in the mixture at thermodynamic equilibrium, and probably play an important role in BNNT synthesis.

Further research on BNNT synthesis and the nitrogen fixation process will be facilitated by the information presented in the supplementary material: this includes rate constants for the dissociative adsorption of $N_2$ on small $B_m$ clusters (m=2-5), as well as rate constants for the reactions leading to folding of $B_{12}N_{12}$ and $B_{24}N_{24}$ monocyclic rings into their corresponding fullborenes. These rate constants can be used for kinetic modeling of fullborene formation. In addition, the standard Gibbs free energy of formation for $B_nN_n$ monocyclic rings, $B_{n+1}N_n$ linear chains, and $B_nN_n$ fullborenes can be found in supplementary materials.

## Acknowledgements

This research was supported in part by the US Office of Fusion Energy Science.

## Supplementary materials

### A1. New mixture components for thermodynamic modeling

Previously, only small $B_mN_n$ species (m<3) were taken into account in past thermodynamic modeling [14]. New species (with m≥3) are considered in our current model, and some of them are presented in Fig. S1.

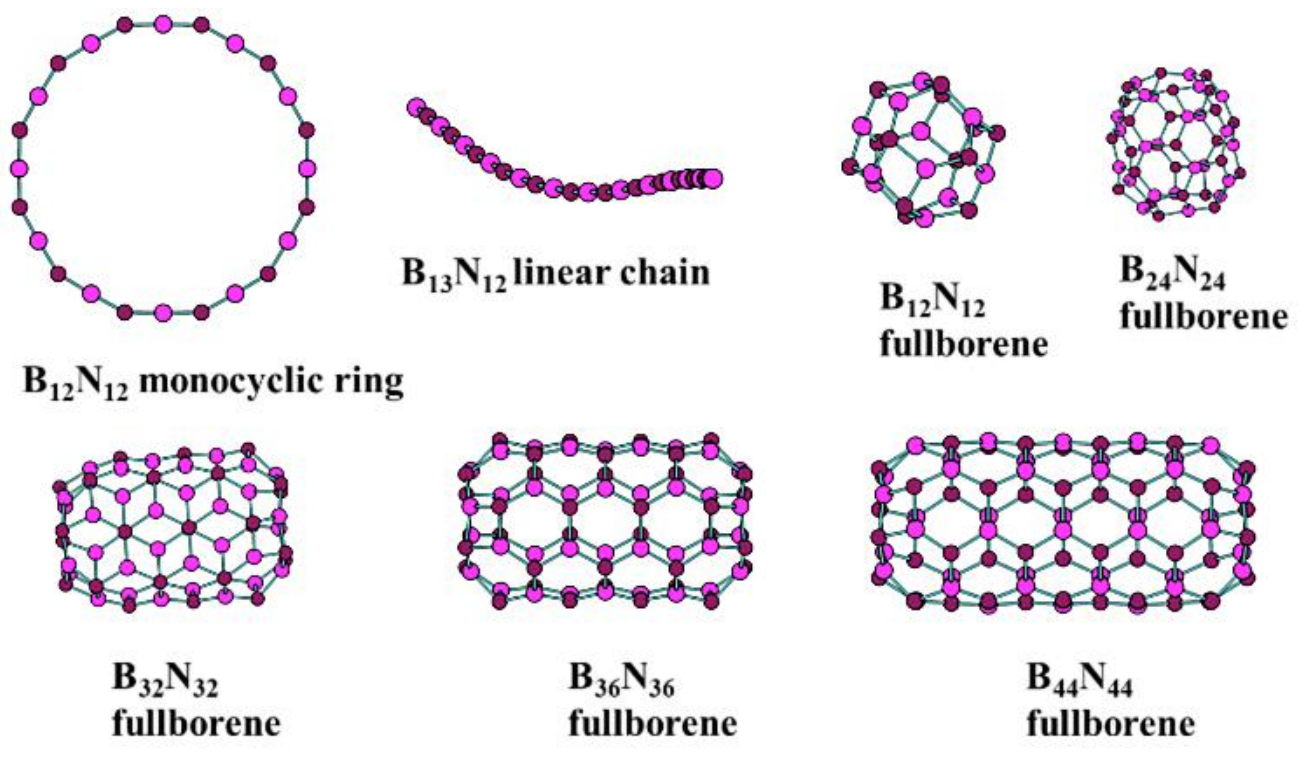

**Figure S1.** New components of an equilibrium reaction mixture, which were considered in our research.

The species under consideration include $B_m$ clusters (m=2-6) and B-N molecules such as $B_nN_n$ monocyclic rings (n=3-12), $B_{n+1}N_n$ linear chains (n=2-12), and $B_nN_n$ fullborenes. In addition, the following hydrogen-containing species were also considered: HBNH, HBNBNH, and $B_3N_3H_6$ (borazine).

### A2. Standard Gibbs free energy of formation

The standard Gibbs free energy of formation ($\Delta_f G^0$) has been calculated for each new species (Fig. S1) using a DFT approach (WB97X-D). The DFT functional and basis set were chosen to better reproduce known values of $\Delta_f G^0(B_2)$ from the NIST database [18] (see Fig. S2). Calculated values of $\Delta_f G^0$ for various species can be found in Table S1-S4.

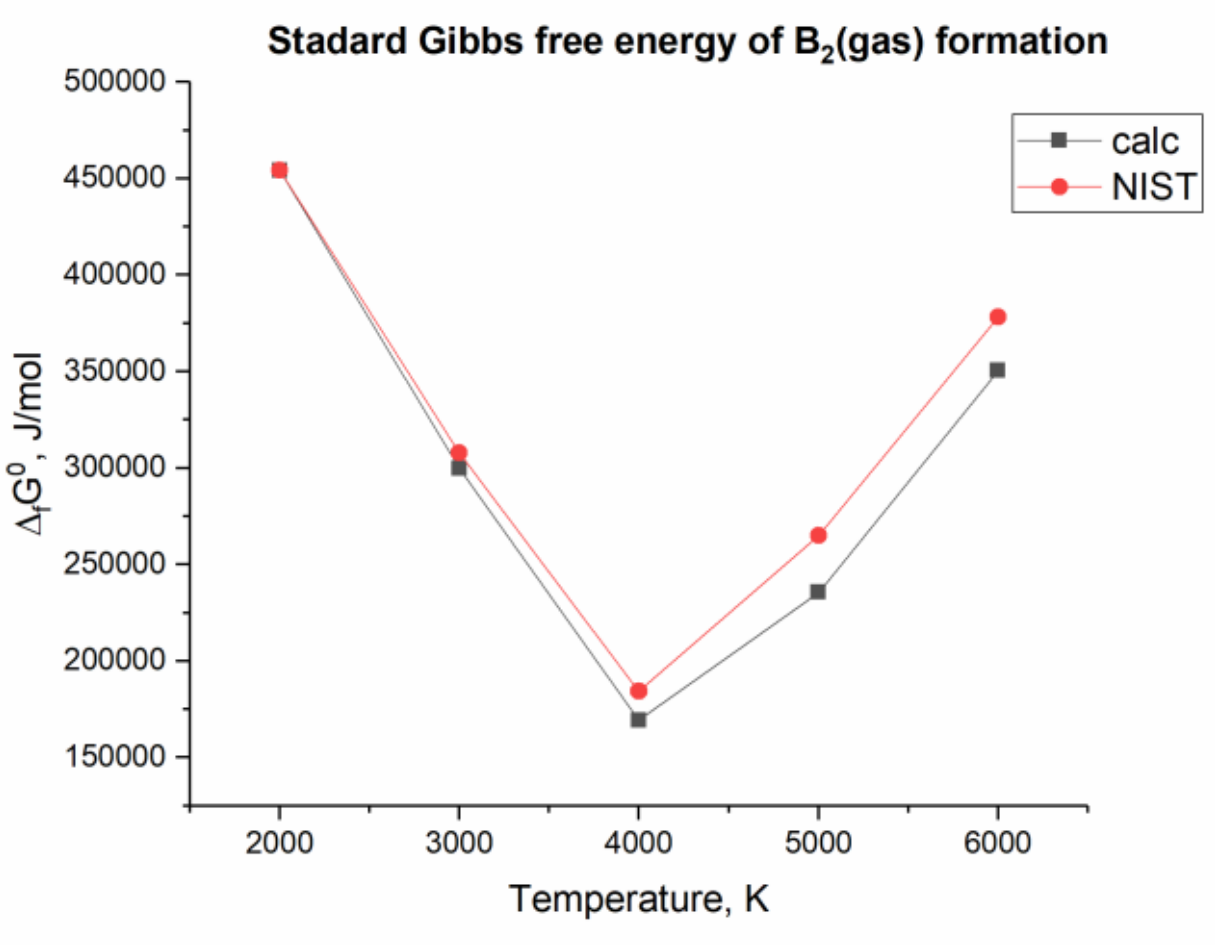

**Figure S2.** Standard Gibbs free energy of $B_2$(gas) formation.

### A3. Self-assembly of small chains

Our DFTB+ modeling is related to the interaction of small $B_3N_3$ cyclic chains with each other at 3000K. Planar collisions of these molecules lead to the formation of one big $B_{12}N_{12}$ chain (Fig. S3a). This result correlates with thermodynamic modeling, which shows that longer B-N chains, both cyclical and linear, are more stable than shorter B-N chains at T < 3000 K. Nonplanar collisions of these rings lead to the formation of $B_{12}N_{12}$ and $B_{24}N_{24}$ fullborenes in the shape of nanocages (see Fig. S3b and S3c). Thus, the smaller B-N rings tend to combine into longer chains and fullborenes. All movies can be found in [38].

We suggest that $B_4N_4$ and $B_5N_4$ will similarly form longer chains and fullborenes by self-assembling.

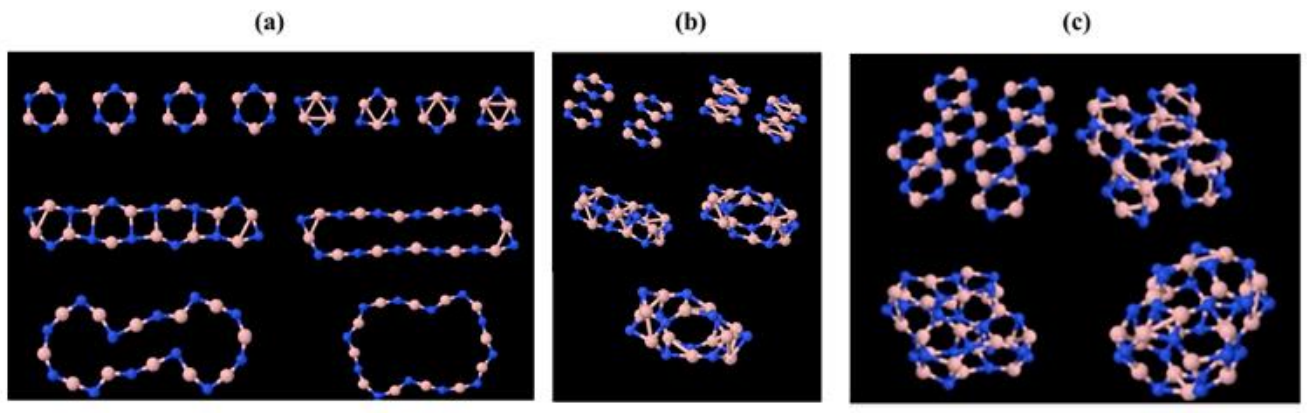

**Figure S3.** Results of DFTB+ simulations: (a) four $B_3N_3$ rings collide in a single plane to form a larger $B_{12}N_{12}$ ring, (b) four $B_3N_3$ rings combine to form a $B_{12}N_{12}$ nanocage, and (c) eight $B_3N_3$ rings combine to form a $B_{24}N_{24}$ nanocage.

### A4. Limitations of DFT and DFTB methods

Here we discuss the limitations of DFT and DFTB methods in the calculation of reaction paths for small species. These limitations occur due to the nature of DFT as a one-determinant method in which only the ground state is taken into account. If a reactant and product have different electronic states, then the transition state cannot be localized by DFT.

As an example, consider the formation of $BN_2$, studied by Martin et al. [26] using the CASSCF (complete active space self-consistent field) method. According this research, there are four relevant states: the ground state of $BN_2$ and three excited states. The ground state structure of $BN_2$ is BNN ($X^2\Pi$), which exists at the same energy level as separated B and $N_2$. The second state – the cyclic $^2A_1$ state – is only slightly higher in energy, and the potential energy surface for interconversion with the linear form possesses a barrier. The $^2B_2$ state is not observed experimentally due to a low barrier towards dissociation to B+$N_2$. This dissociation reaction is exothermic. The last state is the NBN($^2\Pi_g$) state, which has a very high barrier towards interconversion to the cyclic form and subsequent dissociation, and is therefore observed experimentally [26]. An energy diagram of $BN_2$ molecule states can be found in Fig. S4.

Running DFTB+ calculations for the separated B and $N_2$ species, we found that they do not form any bound states of the $BN_2$ molecule. On the other hand, if the initial geometry *is* the linear NBN molecule, then this molecule is stable and does not decompose with time. We did not observe that NBN reverts into a cyclical geometry. This is probably connected to the fact that the transformation of linear NBN to a cyclical geometry proceeds on the excited $^2B_2$ state (see Fig. S4), and this transition cannot be predicted by DFTB.

Indeed, the transition from the $^2\Pi$ state to the $^2B_2$ state through TS1 cannot be evaluated by a one-configuration method such as DFT. This transition requires at least a two-configuration method, which will include both states; only then can TS1 be localized, as its wavefunction is a combination of configurations related to the $^2\Pi$and $^2B_2$ states. Thus, the stability of the linear NBN molecule is probably the result of DFTB being unable to leave the local minimum related to a linear NBN structure towards the cycle structure, since these structures are related to essentially different wavefunctions. Our repeated attempts to localize TS1 with the WB97X-D DFT functional also failed. Generally, the overlap of low-lying excited states with the same multiplicity is an issue for small molecules that have high symmetry. Note that for the BNBN molecule DFT methods can successfully be used, as shown in Fig. 3a.

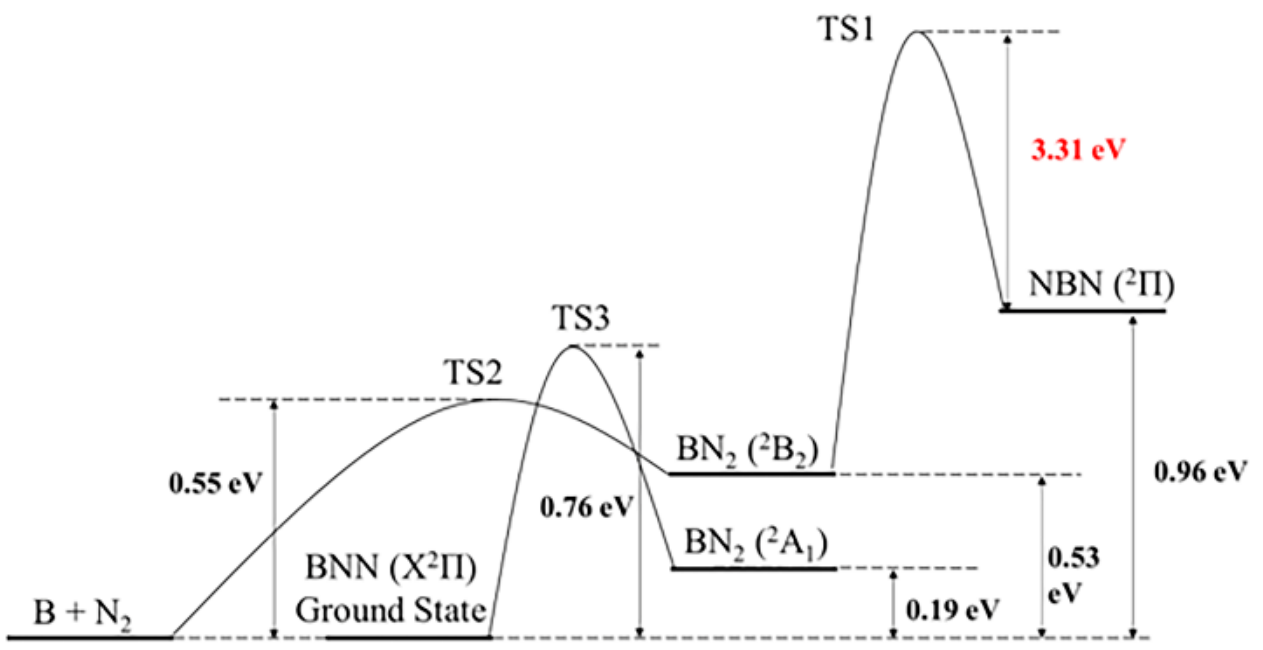

**Figure S4.** Energy diagram of CASSCF calculations performed by Martin et al. [26] to investigate the $BN_2$ molecule formation.

**Table S1.** Calculated standard Gibb free energy of formation (J/mol) of small species.

| T, K | BNBN | $B_3N_2$ | $B_2$ | $B_3$ | $B_4$ | $B_5$ | $B_6$ |
|---|---|---|---|---|---|---|---|
| 1000 | 464251 | 329490 | - | 635992 | 666349 | 715268 | 881623 |
| 2000 | 408482 | 217628 | 454273 | 406567 | 419638 | 441940 | 589541 |
| 2200 | 400535 | 198490 | - | 362691 | 372639 | 389863 | 534012 |
| 2400 | 393404 | 181272 | - | 322628 | 330633 | 343944 | 485815 |
| 2600 | 393050 | 174082 | - | 292792 | 302176 | 314888 | 457799 |
| 2800 | 393045 | 167291 | - | 263538 | 274415 | 286633 | 430695 |
| 3000 | 390277 | 157782 | 299823 | 234822 | 247298 | 259120 | 404436 |
| 3200 | 393980 | 154785 | - | 206608 | 220778 | 232295 | 378958 |
| 3400 | 394878 | 149020 | - | 178863 | 194817 | 206114 | 354214 |
| 3600 | 396039 | 143552 | - | 151558 | 169380 | 180538 | 330157 |
| 3800 | 397448 | 138366 | - | 124669 | 144439 | 155535 | 306749 |
| 4000 | 394521 | 128873 | 169228 | 98173 | 119965 | 131069 | 283953 |
| 4200 | 415000 | 149839 | - | 93118 | 124024 | 142227 | 303872 |
| 4400 | 463341 | 214803 | - | 136747 | 192943 | 234420 | 421003 |
| 4600 | 511684 | 279695 | - | 180416 | 261867 | 326580 | 538064 |
| 4800 | 560029 | 344518 | - | 224123 | 330795 | 418708 | 655057 |
| 5000 | 602287 | 403186 | 235602 | 267866 | 399725 | 510804 | 771985 |
| 6000 | 842505 | 724522 | 350474 | 487078 | 744416 | 970849 | 1355724 |

**Table S2.** Calculated standard Gibb free energy of formation (J/mol) of monocyclic rings.

| T, K | $B_3N_3$ | $B_4N_4$ | $B_5N_5$ | $B_6N_6$ | $B_7N_7$ | $B_8N_8$ | $B_{10}N_{10}$ | $B_{12}N_{12}$ | $B_{24}N_{24}$ |
|---|---|---|---|---|---|---|---|---|---|
| 1000 | 146185 | -47445 | -174285 | -266139 | -338214 | -405134 | -537716 | -659192 | -1264888 |
| 2000 | 168424 | 10210 | -59886 | -120574 | -147984 | -178390 | -247860 | -304293 | -534435 |
| 2200 | 176718 | 26508 | -31301 | -84870 | -102384 | -124565 | -179576 | -221163 | -365123 |
| 2400 | 185980 | 44001 | -1290 | -47519 | -54903 | -68632 | -108729 | -135016 | -190062 |
| 2600 | 205170 | 74644 | 45093 | 9423 | 15394 | 13339 | -5402 | -9946 | 62581 |
| 2800 | 224666 | 105612 | 91822 | 66727 | 86078 | 95717 | 98371 | 115608 | 315950 |
| 3000 | 239812 | 130705 | 131152 | 115104 | 146312 | 166119 | 187117 | 223084 | 532934 |
| 3200 | 264475 | 168414 | 186200 | 182296 | 228465 | 261540 | 307082 | 367979 | 824542 |
| 3400 | 284749 | 200205 | 233800 | 240509 | 300111 | 344924 | 411950 | 494718 | 1079635 |
| 3600 | 305249 | 232233 | 281650 | 298982 | 372030 | 428593 | 517125 | 621786 | 1335199 |
| 3800 | 325961 | 264485 | 329734 | 357699 | 444204 | 512528 | 622589 | 749164 | 1591200 |
| 4000 | 340017 | 287804 | 366610 | 402931 | 500620 | 578429 | 705465 | 849403 | 1792753 |
| 4200 | 389039 | 357690 | 461656 | 517932 | 638407 | 737301 | 904516 | 1089018 | 2272896 |
| 4400 | 479714 | 483062 | 626020 | 716084 | 873178 | 1006992 | 1242049 | 1494781 | 3085179 |
| 4600 | 570262 | 608214 | 790073 | 913830 | 1107453 | 1276093 | 1578810 | 1899587 | 3895398 |
| 4800 | 660687 | 733155 | 953826 | 1111189 | 1341252 | 1544630 | 1914829 | 2303474 | 4703639 |
| 5000 | 741862 | 845717 | 1102070 | 1289910 | 1553284 | 1788271 | 2219693 | 2669948 | 5436918 |
| 6000 | 1189471 | 1463547 | 1911607 | 2265106 | 2708335 | 3114654 | 3878767 | 4663644 | 9424083 |

**Table S3.** Calculated standard Gibb free energy of formation (J/mol) of $B_{n+1}N_n$ linear chains.

| T, K | $B_5N_4$ (linear) | $B_5N_4$ (cycle) | $B_7N_6$ | $B_8N_7$ | $B_9N_8$ | $B_{10}N_9$ | $B_{11}N_{10}$ | $B_{12}N_{11}$ | $B_{13}N_{12}$ |
|---|---|---|---|---|---|---|---|---|---|
| 1000 | 234155 | 44749 | 48092 | -20027 | -96887 | -172073 | -245523 | -310933 | -382740 |
| 2000 | 185936 | 58761 | 37178 | -4846 | -63518 | -118019 | -168449 | -201949 | -247966 |
| 2200 | 181484 | 66595 | 41895 | 6005 | -48116 | -97565 | -142476 | -168681 | -208625 |
| 2400 | 179426 | 76792 | 49436 | 19907 | -29436 | -73607 | -112774 | -131456 | -165101 |
| 2600 | 193857 | 103455 | 79890 | 59942 | 18598 | -17076 | -47277 | -55216 | -79342 |
| 2800 | 208739 | 130550 | 110820 | 100473 | 67145 | 39988 | 18772 | 21596 | 7009 |
| 3000 | 217863 | 151870 | 132928 | 130655 | 103818 | 83652 | 69894 | 81954 | 75379 |
| 3200 | 239708 | 185899 | 173947 | 182848 | 165602 | 155526 | 152327 | 176723 | 181260 |
| 3400 | 255733 | 214096 | 206077 | 224622 | 215436 | 213921 | 219750 | 254951 | 269070 |
| 3600 | 272086 | 242611 | 238550 | 266751 | 265636 | 272693 | 287561 | 333579 | 357291 |
| 3800 | 288746 | 271426 | 271346 | 309213 | 316180 | 331818 | 355736 | 412582 | 445898 |
| 4000 | 296554 | 291382 | 290732 | 335989 | 348761 | 370706 | 401397 | 466794 | 507437 |
| 4200 | 358026 | 364996 | 386980 | 451228 | 481406 | 521257 | 570324 | 655872 | 715444 |
| 4400 | 491164 | 510271 | 582556 | 679626 | 741044 | 812634 | 893907 | 1013440 | 1105773 |
| 4600 | 624049 | 655288 | 777690 | 907492 | 1000056 | 1103292 | 1216681 | 1370105 | 1495107 |
| 4800 | 756692 | 800059 | 972402 | 1134846 | 1258468 | 1393263 | 1538677 | 1725906 | 1883487 |
| 5000 | 876924 | 932416 | 1148443 | 1340399 | 1491950 | 1655174 | 1829485 | 2047388 | 2234420 |
| 6000 | 1532681 | 1648756 | 2109782 | 2462590 | 2767069 | 3085724 | 3417805 | 3802365 | 4149936 |

**Table S4.** Calculated standard Gibb free energy of formation (J/mol) of $B_nN_n$ fullborenes.

| T, K | $B_{12}N_{12}$ | $B_{24}N_{24}$ | $B_{32}N_{32}$ | $B_{36}N_{36}$ | $B_{44}N_{44}$ |
|---|---|---|---|---|---|
| 1000 | -981850 | -2372662 | -3551630 | -4151162 | -5354925 |
| 2000 | -238194 | -684688 | -1226831 | -1507786 | -2074428 |
| 2200 | -77556 | -324049 | -731411 | -944947 | -1376780 |
| 2400 | 86050 | 42287 | -228476 | -373685 | -668898 |
| 2600 | 288541 | 486162 | 377765 | 313766 | 180940 |
| 2800 | 491485 | 930725 | 984851 | 1002141 | 1031853 |
| 3000 | 676324 | 1338873 | 1543313 | 1635786 | 1815826 |
| 3200 | 898561 | 1821618 | 2201172 | 2381229 | 2736394 |
| 3400 | 1102623 | 2267827 | 2810253 | 3071772 | 3589819 |
| 3600 | 1307000 | 2714488 | 3419876 | 3762902 | 4443921 |
| 3800 | 1511673 | 3161570 | 4030004 | 4454579 | 5298651 |
| 4000 | 1689197 | 3554190 | 4567462 | 5064481 | 6053398 |
| 4200 | 2006086 | 4225387 | 5476303 | 6092170 | 7318737 |
| 4400 | 2489115 | 5228714 | 6827934 | 7617978 | 9192855 |
| 4600 | 2971179 | 6229967 | 8176753 | 9140605 | 11063051 |
| 4800 | 3452319 | 7229235 | 9522878 | 10660183 | 12929491 |
| 5000 | 3896040 | 8153532 | 10768998 | 12067240 | 14658374 |
| 6000 | 6275902 | 13095706 | 17426570 | 19582638 | 23889182 |

**Table S5.** Calculated rate constants for the reactions of $N_2$ fixations and monocyclic ring transformation into the fullborene. Reactions paths related to these constants can be found in Fig. 3, S4 and S5. "f" and "r" symbols mean forward and reverse reactions.

| No. | Reaction | A(f) | n(f) | Ea(f), eV | A(r) | n(r) | Ea(r), eV |
|---|---|---|---|---|---|---|---|
| | $B_2 + N_2 \leftrightarrow$ BNBN | | | | | | |
| 1 | $B_2 + N_2 \leftrightarrow$ BBNN | $3.26 \cdot 10^{-18}$ $m^3$/s | 2.0 | 0.23 | $2.91 \cdot 10^{15}$ 1/s | 0 | 1.50 |
| 2 | BBNN $\leftrightarrow$ $BBN_2$ | $8.46 \cdot 10^{13}$ 1/s | -0.25 | 1.10 | $1.45 \cdot 10^{13}$ 1/s | 0.15 | 1.08 |
| 3 | $BBN_2 \leftrightarrow B_2N_2$ | $4.89 \cdot 10^{12}$ 1/s | 0.1 | 0.76 | $1.40 \cdot 10^{13}$ 1/s | 0.1 | 0.79 |
| 4 | $B_2N_2 \leftrightarrow$ BNBN | $1.45 \cdot 10^{13}$ 1/s | 0.1 | 0.45 | $8.57 \cdot 10^{12}$ 1/s | -0.3 | 2.49 |
| | $B_3 + N_2 \leftrightarrow B_3N_2$ | | | | | | |
| 1 | $B_3 + N_2 \leftrightarrow B_3N_2$ (1) | $1.35 \cdot 10^{-18}$ $m^3$/s | 2.4 | 0.16 | $7.79 \cdot 10^{14}$ 1/s | 0.35 | 0.72 |
| 2 | $B_3N_2$ (1) $\leftrightarrow B_3N_2$ (2) | $3.02 \cdot 10^{12}$ 1/s | 0.15 | 1.57 | $1.10 \cdot 10^{13}$ 1/s | 0.07 | 0.08 |
| 3 | $B_3N_2$ (2) $\leftrightarrow B_3N_2$ (3) | $6.38 \cdot 10^{12}$ 1/s | 0.05 | 0.12 | $1.56 \cdot 10^{13}$ 1/s | 0.1 | 0.83 |
| 4 | $B_3N_2$ (3) $\leftrightarrow B_3N_2$ | $2.84 \cdot 10^{13}$ 1/s | 0.1 | 0.29 | $1.29 \cdot 10^{13}$ 1/s | 0.25 | 4.81 |
| | $B_4 + 2N_2 \leftrightarrow B_4N_4$ | | | | | | |
| 1 | $B_4 + N_2 \leftrightarrow B_4N_2$ (1) | $7.79 \cdot 10^{-19}$ $m^3$/s | 2.37 | 0.09 | $1.40 \cdot 10^{15}$ 1/s | 0.29 | 0.76 |
| 2 | $B_4N_2$ (1) $\leftrightarrow B_4N_2$ (2) | $2.86 \cdot 10^{12}$ 1/s | 0.12 | 1.17 | $1.58 \cdot 10^{13}$ 1/s | 0.05 | 0.19 |
| 3 | $B_4N_2$ (2) $\leftrightarrow B_4N_2$ | $1.55 \cdot 10^{14}$ 1/s | 0.25 | 0.89 | $1.16 \cdot 10^{14}$ 1/s | 0.52 | 5.56 |
| 4 | $B_4N_2 + N_2 \leftrightarrow B_4N_4$ (1) | $2.03 \cdot 10^{-19}$ $m^3$/s | 2.45 | 0.40 | $9.48 \cdot 10^{14}$ 1/s | 0.45 | 1.83 |
| 5 | $B_4N_4$ (1) $\leftrightarrow B_4N_4$ (2) | $6.47 \cdot 10^{13}$ 1/s | 0.35 | 1.62 | $1.28 \cdot 10^{14}$ 1/s | 0.25 | 1.09 |
| 6 | $B_4N_4$ (2) $\leftrightarrow B_4N_4$ (3) | $3.40 \cdot 10^{12}$ 1/s | 0.13 | 1.52 | $6.22 \cdot 10^{11}$ 1/s | 0.06 | 1.35 |
| 7 | $B_4N_4$ (3) $\leftrightarrow B_4N_4$ | $3.19 \cdot 10^{12}$ 1/s | 0.08 | 0.24 | $4.45 \cdot 10^{13}$ 1/s | 0.4 | 5.78 |
| | $B_5 + 2N_2 \leftrightarrow B_5N_4$ | | | | | | |
| 1 | $B_5 + N_2 \leftrightarrow B_5N_2$ (1) | $8.20 \cdot 10^{-19}$ $m^3$/s | 2.4 | 0.13 | $2.05 \cdot 10^{14}$ 1/s | 0.3 | 1.05 |
| 2 | $B_5N_2$ (1) $\leftrightarrow B_5N_2$ (2) | $8.14 \cdot 10^{12}$ 1/s | 0.2 | 1.09 | $5.51 \cdot 10^{12}$ 1/s | 0.05 | 0.04 |
| 3 | $B_5N_2$ (3) $\leftrightarrow B_5N_2$ (4) | $9.46 \cdot 10^{11}$ 1/s | -0.05 | 0.71 | $1.08 \cdot 10^{13}$ 1/s | 0.1 | 1.08 |
| 4 | $B_5N_2$ (4) $\leftrightarrow B_5N_2$ | $2.83 \cdot 10^{13}$ 1/s | 0.2 | 0.80 | $1.65 \cdot 10^{13}$ 1/s | 0.4 | 4.94 |
| 5 | $B_5N_2 + N_2 \leftrightarrow B_5N_4$ (1) | $8.66 \cdot 10^{-18}$ $m^3$/s | 2.4 | 0 | $2.43 \cdot 10^{16}$ 1/s | 0.4 | 0.93 |
| 6 | $B_5N_4$ (1) $\leftrightarrow B_5N_4$ (2) | $5.06 \cdot 10^{12}$ 1/s | 0.2 | 1.71 | $1.99 \cdot 10^{12}$ 1/s | 0 | 0 |
| 7 | $B_5N_4$ (2) $\leftrightarrow B_5N_4$ (3) | $1.22 \cdot 10^{12}$ 1/s | 0.09 | 0.17 | $4.20 \cdot 10^{12}$ 1/s | 0.15 | 0.32 |
| 8 | $B_5N_4$ (3) $\leftrightarrow B_5N_4$ | $1.83 \cdot 10^{13}$ 1/s | 0.1 | 0.40 | $3.87 \cdot 10^{13}$ 1/s | 0.35 | 5.98 |
| | monocyclic ring $\leftrightarrow$ fullborene ($B_{12}N_{12}$) | | | | | | |
| 1 | $B_{12}N_{12}$ monocyclic ring $\leftrightarrow B_{12}N_{12}$-(1) | $8.88 \cdot 10^{10}$ 1/s | -0.05 | 1.77 | $4.17 \cdot 10^{13}$ 1/s | 0.25 | 0.18 |
| 2 | $B_{12}N_{12}$-(1) $\leftrightarrow B_{12}N_{12}$-(2) | $3.82 \cdot 10^{11}$ 1/s | 0 | 1.93 | $1.18 \cdot 10^{15}$ 1/s | 0.35 | 0.65 |
| 3 | $B_{12}N_{12}$-(2) $\leftrightarrow B_{12}N_{12}$-(3) | $1.25 \cdot 10^{12}$ 1/s | 0.07 | 0.36 | $1.25 \cdot 10^{14}$ 1/s | 0.25 | 0.16 |
| 4 | $B_{12}N_{12}$-(3) $\leftrightarrow B_{12}N_{12}$-(4) | $1.73 \cdot 10^{11}$ 1/s | -0.15 | 0.73 | $3.20 \cdot 10^{14}$ 1/s | 0.5 | 2.44 |
| 5 | $B_{12}N_{12}$-(3) $\leftrightarrow B_{12}N_{12}$ fullborene | $2.30 \cdot 10^{11}$ 1/s | -0.04 | 0.02 | $2.32 \cdot 10^{16}$ 1/s | 1.0 | 5.86 |
| | monocyclic ring $\leftrightarrow$ fullborene ($B_{24}N_{24}$) | | | | | | |
| 1 | $B_{24}N_{24}$ monocyclic ring $\leftrightarrow B_{24}N_{24}$-(1) | $2.45 \cdot 10^{10}$ 1/s | -0.05 | 0.79 | $8.68 \cdot 10^{14}$ 1/s | 0.35 | 1.37 |
| 2 | $B_{24}N_{24}$-(1) $\leftrightarrow B_{24}N_{24}$-(2) | $4.69 \cdot 10^{10}$ 1/s | 0.13 | 0.94 | $7.52 \cdot 10^{12}$ 1/s | 0.10 | 0.05 |
| 3 | $B_{24}N_{24}$-(2) $\leftrightarrow B_{24}N_{24}$-(3) | $5.87 \cdot 10^{12}$ 1/s | 0.10 | 0.05 | $2.94 \cdot 10^{12}$ 1/s | 0.10 | 0.94 |
| 4 | $B_{24}N_{24}$-(3) $\leftrightarrow B_{24}N_{24}$-(4) | $2.83 \cdot 10^{8}$ 1/s | -0.15 | 3.06 | $1.02 \cdot 10^{15}$ 1/s | 0.50 | 2.86 |
| 5 | $B_{24}N_{24}$-(4) $\leftrightarrow B_{24}N_{24}$-(5) | $1.36 \cdot 10^{13}$ 1/s | 0.12 | 0.16 | $2.98 \cdot 10^{12}$ 1/s | 0.07 | 0.32 |
| 6 | $B_{24}N_{24}$-(5) $\leftrightarrow B_{24}N_{24}$-(6) | $3.50 \cdot 10^{12}$ 1/s | 0.05 | 0.17 | $2.47 \cdot 10^{13}$ 1/s | 0.15 | 0.26 |
| 7 | $B_{24}N_{24}$-(6) $\leftrightarrow B_{24}N_{24}$-(7) | $7.43 \cdot 10^{11}$ 1/s | 0.04 | 0.56 | $1.55 \cdot 10^{13}$ 1/s | 0.12 | 0.06 |
| 8 | $B_{24}N_{24}$-(7) $\leftrightarrow B_{24}N_{24}$-(8) | $7.25 \cdot 10^{11}$ 1/s | -0.05 | 1.22 | $1.08 \cdot 10^{16}$ 1/s | 0.60 | 2.74 |
| 9 | $B_{24}N_{24}$-(8) $\leftrightarrow B_{24}N_{24}$-(9) | $7.60 \cdot 10^{11}$ 1/s | 0.15 | 2.30 | $3.80 \cdot 10^{15}$ 1/s | 0.60 | 3.89 |
| 10 | $B_{24}N_{24}$-(9) $\leftrightarrow B_{24}N_{24}$-(10) | $2.52 \cdot 10^{11}$ 1/s | -0.02 | 0.52 | $7.53 \cdot 10^{13}$ 1/s | 0.20 | 0.77 |
| 11 | $B_{24}N_{24}$-(10) $\leftrightarrow B_{24}N_{24}$(11) | $1.62 \cdot 10^{11}$ 1/s | 0 | 0.76 | $1.38 \cdot 10^{13}$ 1/s | 0.12 | 0.07 |
| 12 | $B_{24}N_{24}$-(11) $\leftrightarrow B_{24}N_{24}$(12) | $1.01 \cdot 10^{12}$ 1/s | -0.01 | 0.67 | $3.62 \cdot 10^{15}$ 1/s | 0.60 | 3.02 |
| 13 | $B_{24}N_{24}$-(12) $\leftrightarrow B_{24}N_{24}$(13) | $5.52 \cdot 10^{11}$ 1/s | 0.01 | 0.82 | $7.67 \cdot 10^{12}$ 1/s | 0.10 | 0.21 |
| 14 | $B_{24}N_{24}$-(13) $\leftrightarrow B_{24}N_{24}$(14) | $9.30 \cdot 10^{11}$ 1/s | -0.03 | 0.11 | $1.43 \cdot 10^{13}$ 1/s | 0.15 | 1.31 |
| 15 | $B_{24}N_{24}$-(14) $\leftrightarrow B_{24}N_{24}$(15) | $1.41 \cdot 10^{12}$ 1/s | 0.01 | 0.33 | $3.36 \cdot 10^{10}$ 1/s | 1.30 | 10.51 |
| 16 | $B_{24}N_{24}$-(15) $\leftrightarrow B_{24}N_{24}$ fullborene | $3.09 \cdot 10^{12}$ 1/s | 0.02 | 0.21 | $5.77 \cdot 10^{15}$ 1/s | 0.80 | 5.75 |